\documentclass[pre,twocolumn,superscriptaddress,floatfix]{revtex4}
\usepackage{epsfig}
\usepackage{subfigure}
\usepackage{graphicx}
\usepackage{amssymb}

\begin{document}

\title{
Simulation of a two-dimensional model for colloids in a uniaxial electric field}

\author{Ahmad M. Almudallal}
\affiliation{Department of Physics and Physical Oceanography,
Memorial University of Newfoundland, St. John's, NL, A1B 3X7, Canada}

\author{Ivan Saika-Voivod}
\affiliation{Department of Physics and Physical Oceanography,
Memorial University of Newfoundland, St. John's, NL, A1B 3X7, Canada}

\date{\today}

\begin{abstract}

We perform Monte Carlo simulations of a simplified two-dimensional model for colloidal hard spheres in 
an external uniaxial AC electric field.  Experimentally, the external field induces dipole moments in the colloidal particles,
which in turn form chains.  We therefore approximate the system as composed of well formed chains of dipolar hard spheres
of a uniform length.  
The dipolar interaction between colloidal spheres gives rise to an effective interaction 
between the chains, which we treat as disks in a plane, that includes a short range attraction and long range 
repulsion.
Hence, the system favors finite clustering over bulk phase separation and indeed we observe 
at low temperature and density that the system does form a cluster phase.
As density increases, percolation is accompanied by a pressure anomaly.  
The percolated phase, despite being composed of connected, locally crystalline domains, does not bear the typical signatures of a hexatic phase.
At very low densities, we find no indication of a ``void phase'' with a cellular structure seen recently in experiments.

\end{abstract}

\maketitle

\section{INTRODUCTION}

Colloidal suspensions, small particles dispersed within a second medium, are common in
everyday life with examples ranging from toothpaste to paint, from milk to quicksand
\cite{fennell-evans,russel,robert-hunter,richard}.
An interesting class of suspension is the electrorheological (ER) fluid
\cite{stangroom,woestman,halsey,klingenberg,tao3,chen,tian}, where the liquid
medium and the colloidal particles have different dielectric constants.  
An externally applied electric field polarizes the particles, and hence 
the system can be thought of as a system of dipolar particles,
the moments of which are aligned along the field axis.

ER fluids and their phase behavior have been studied experimentally~\cite{wang,anand-alfons,anand-thijssen,ay2002,anil,amit}, 
theoretically~\cite{halsey,tao3} and using computer simulation~\cite{dijkstra,tao2,tao1,james1,james2}.
In Ref~\cite{dijkstra}, Hynninen and Dijkstra used MC simulation to study the phase behavior of hard 
and soft dipolar spheres over a broad range of packing fraction and external field strength.  The morphologies they saw
match those seen experimentally.  What is observed both in simulation and experiment is the formation of chains
of particles that in turn form sheets and 
body centered tetragonal ($bct$) clusters, which have been interpreted at low temperature $T$ to be the crystal phase coexisting with a gas.  

Also very interesting are the experimental reports of a ``void phase'', where at very low particle concentrations and sufficient external field strength a cell structure appears with relatively colloid rich walls forming the border between colloid poor voids.  The structure has been reported both in the granular~\cite{anil} and Brownian~\cite{amit} regimes, where it appears to be a stable structure.  A similar but transient structure appears as a result of a Rayleigh-Taylor-like hydrodynamic instability during sedimentation~\cite{lowen2009}.

Both the formation of $bct$ crystallites and the void phase occur at higher external field strengths, in the regime where particles 
form long chains along the field axis.  Our goal is to gain a simpler understanding of the structures formed in the system, and so 
we present a simplified model in which the system is represented by well formed (rigid) chains of dipolar hard spheres of uniform length.  In this case, the system is essentially two dimensional, with disks representing chains as viewed down the field axis.  While this is perhaps not physically realistic, it should capture the essential features of the real system at high field strength, at least in terms of packing and dipolar effects.

Our desire to use this simplified model stems from several factors.  First, we believe that the finite extent of experimental systems plays a significant role in the phase behavior seen.  Second, the interesting morphologies observed experimentally are seen at high fields, in the regime where the system is largely composed of chains.  It would be convenient to think of these morphologies arising from interacting chains of dipolar hard spheres, but are these morphologies in fact recovered from interacting chains?  If not, what additional ingredients must be present to recover them?

We believe that a set of chains of uniform length provide a reasonable starting point for a coarse-grained description of the experimental systems, in which this length seems to be between 50 and 100 or so spheres.  We note that simply using hard rods with a uniform dipole moment density would not result in attraction between chains at short distances.  Once the physics of this simple system is understood, progressive detail can be added, such as the effect of chain length, chain length polydispersity (to better simulate a sediment), fluctuations in chain shape etc.

The main feature that emerges when reducing the system to one of  chains, is that the effective interaction
between chains as a function of their separation, while depending on chain length, is generally repulsive at long range and attractive at short range.  This immediately suggests the possibility of clusters that are finite in size that reduce the likelihood of bulk phase separation.

The model embeds the finite extent of the system along the field axis within the chain-chain interactions.  It thus provides a convenient way of approaching the bulk three dimensional limit without progressively increasing the number of particles in the simulation.  In this work, as a starting point and for simplicity, we study a system of chains each composed of $50$ particles, motived for example by Ref.~\cite{ay2002}.

This paper is organized as follows.  In Section II, we present the details of our two dimensional model and the resulting effective potential.  We also discuss the energetics of clustering and provide details on our Monte Carlo simulations, as well as on the descriptive quantities we calculate.  In Section III we present our results, and in Section IV we discuss them, especially with regard to
phase diagrams pertinent to our system.  Finally, we give our Conclusions and thoughts for future work.

\section{Model and simulations}

\subsection{Interacting chains of dipolar particles}

\begin{figure}[htp]
\centering
\subfigure{
\includegraphics[clip=true, trim=0 0 0 0, height=3.75cm, width=3.30cm]{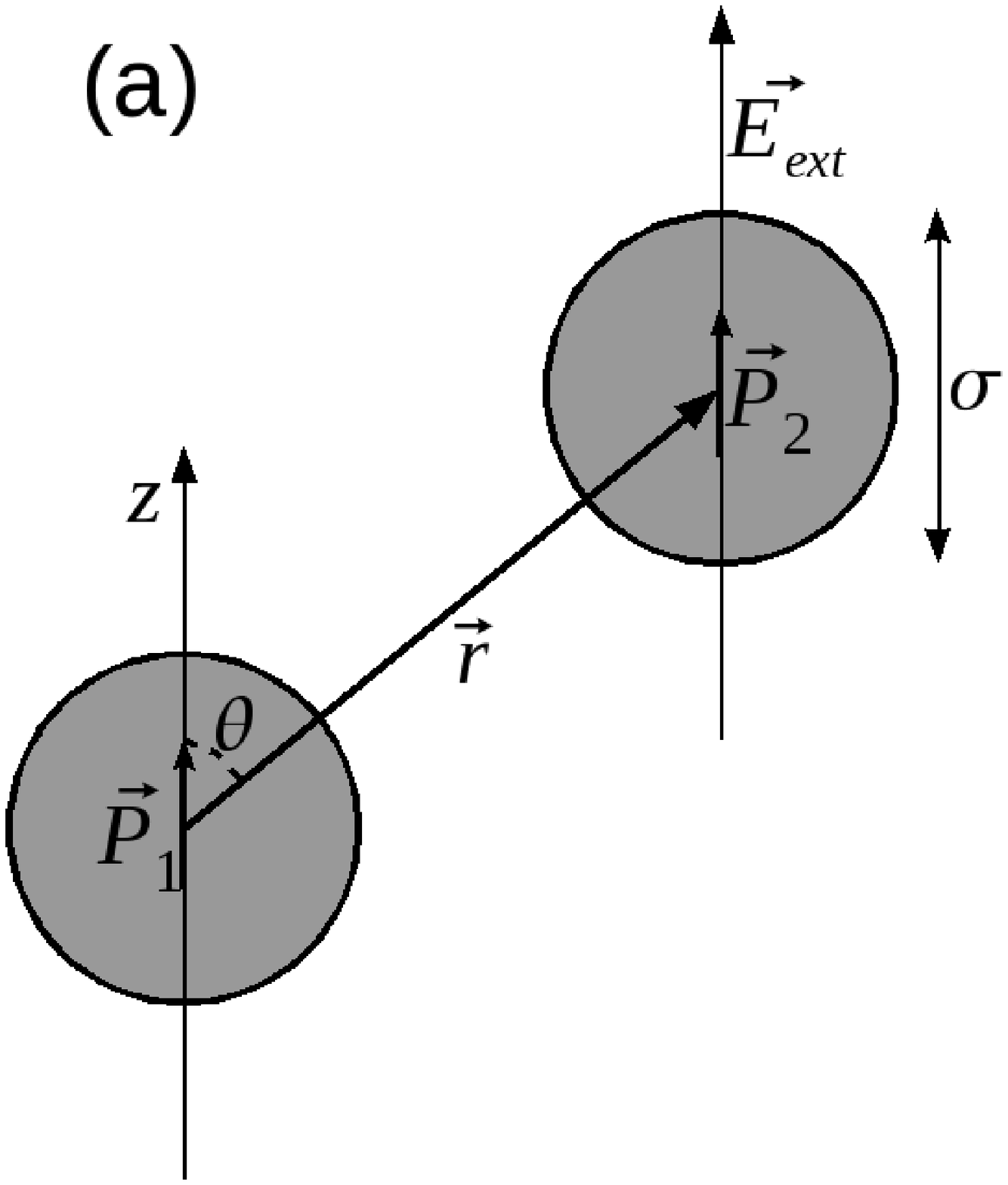}
\label{dipole-angle}
}
\subfigure{
\includegraphics[clip=true, trim=0 0 0 0, height=3.65cm, width=2.03cm]{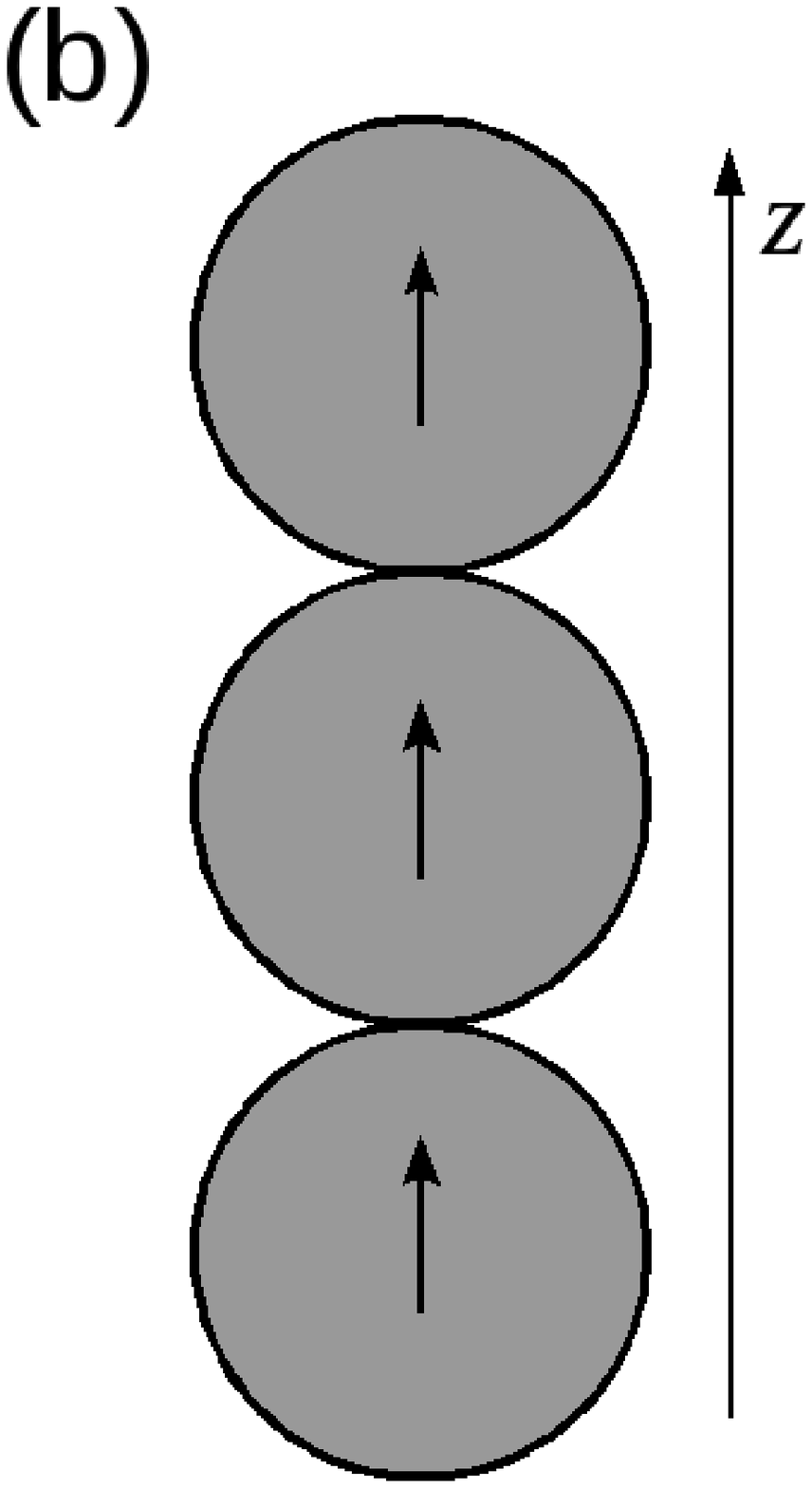}
\label{dipole-angle-long}
}
\caption{\subref{dipole-angle} The interaction between two particles of diameter 
$\sigma$ and parallel dipole moments, $\vec{P}_1$ and $\vec{P}_2$, separated by a distance $|\vec{r}|$. 
$\theta$ is the angle formed between the dipole moment and the separation vector $\vec{r}$. 
The external electric field $\vec{E}_{ext}$ is parallel to the $z-$axis. 
\subref{dipole-angle-long} The preferable configuration of the dipoles under a relatively strong electric field.}
\end{figure}

Our approach to modeling a system of colloidal particles in a uniaxial electric field begins with the assumption that the field induces point dipole moments within otherwise hard spheres of diameter $\sigma$.  As the dipole moments are always aligned, the effective pair interaction becomes,
\begin{equation}\label{dipole-energy}
U(r, \theta) = A_D \left( \frac{\sigma}{r} \right)^3 \left( 1 - 3 \cos^2\theta \right),
\end{equation}
where,
\begin{equation}
A_D = \frac{\pi\sigma^3 \alpha^2 \epsilon_S |E_{loc}|^2}{16},
\label{dipolar-strength}
\end{equation}
is the energy scale of the dipolar interaction, $\alpha = (\epsilon_P-\epsilon_S)/(\epsilon_P+2\epsilon_S)$ is the polarization, $\epsilon_S$ and $\epsilon_P$ are the dielectric constants of the solution and colloids, respectively, 
$|\vec{E}_{loc}| = |\vec{E}_{ext}| + |\vec{E}_{dip}|$ is the magnitude of the local electric field  with a dominant contribution from the applied external field and a negligible contribution from the field from the other induced dipoles.  As shown in Fig.~\ref{dipole-angle}, $\theta$ is the angle formed between the displacement vector between two particles and the field axis, which we choose to be along the $z$ aixs.

At sufficiently small reduced temperature $T^*=k_B T/A_D$, the dipolar spheres tend to align and form chains along the field axis as illustrated in Fig.~\ref{dipole-angle-long}.  Thus, we will model the system as one comprising chains, each composed of $L$ dipolar spheres.  Colloids within a chain are assumed to be fixed with respect to each other, while each chain as a whole may interact with other chains.

\begin{figure}[htp]
\centering
\subfigure{
\includegraphics[clip=true, trim=0 0 0 0, height=3.95cm, width=2.57cm]{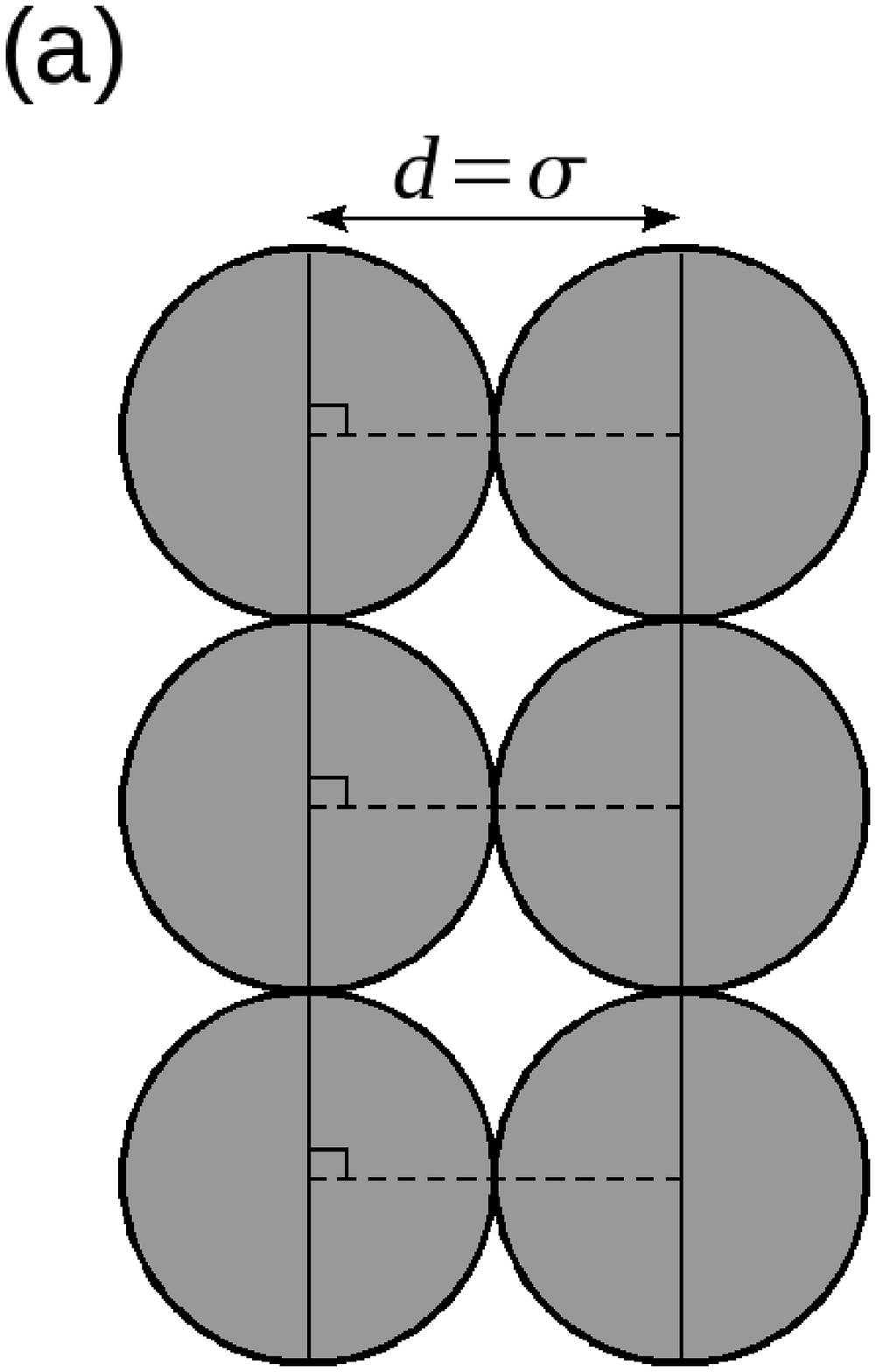}
\label{stacked-closest}
}
\subfigure{
\includegraphics[clip=true, trim=0 0 0 0, height=4.14cm, width=2.54cm]{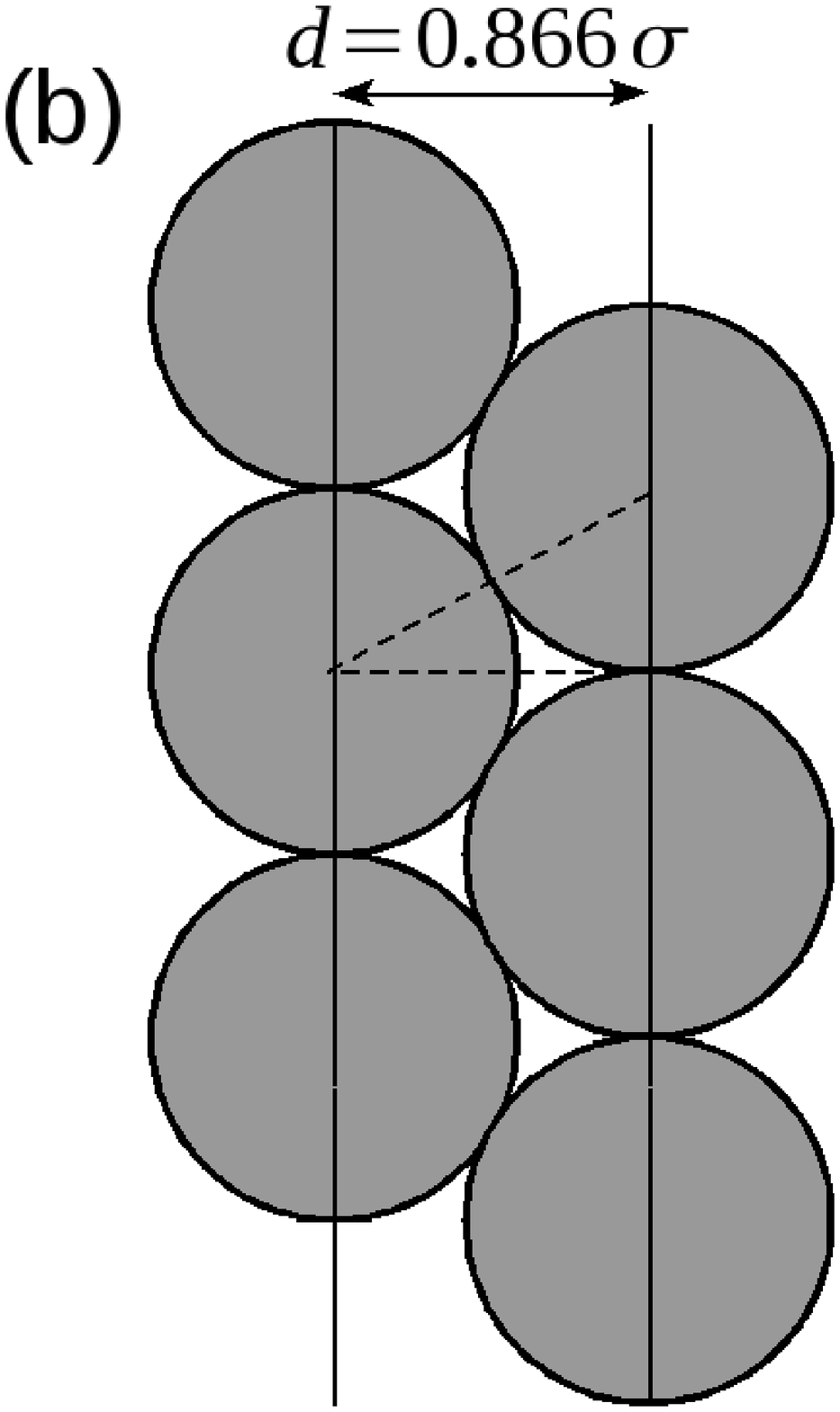}
\label{staggered-closest}
}
\caption{Two chains interacting at closest distance through (a) the stacked interaction and (b) the staggered interaction.}
\end{figure}

Within this approximation, the crystal phase available to the chains is the body centred tetragonal ($bct$) structure, within which chains are either in a ``stacked'' configuration, as in Fig.~\ref{stacked-closest}, or in a ``staggered'' configuration, as in Fig.~\ref{staggered-closest}.  While the staggered configuration is energetically preferred, and indeed is attractive at small chain axis to axis separation $d$, basic geometrical frustration prevents all configurations from being staggered.

\begin{figure}[ht]
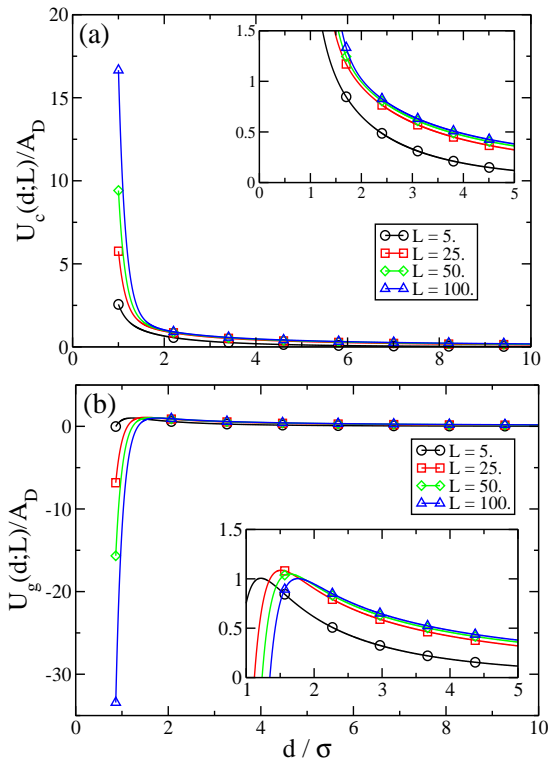

\centering
\subfigure{
\includegraphics[clip=true, trim=0 0 0 0, height=4.8cm, width=7.0cm]{fig3a}
\label{stacked-simul}
}
\subfigure{
\includegraphics[clip=true, trim=0 0 0 0, height=5.0cm, width=7.2cm]{fig3b}
\label{staggered-simul}
}
\caption{(Color online) Chain-chain interaction energy  
as a function of distance ($d$) for different chain lengths ($L$), for 
\subref{stacked-simul} the stacked dipolar interaction, 
exhibiting weak repulsion at large $d$, and strong repulsion at small $d$; and  
\subref{staggered-simul} the staggered dipolar interaction, 
exhibiting weak repulsion at large $d$, and strong attraction at small $d$. 
The inset in each figure shows a close-up of the interaction at short distances.}
\end{figure}

For the stacked geometry, the potential energy as a function of $d$ between two chains is given through Eq.~\ref{dipole-energy} by summing all interactions between pairs of colloids in different chains,
\begin{equation}
\frac{U_{c}(d;L)}{ \sigma^3 A_D}=  \sum_{n=1}^{L}{\sum_{m=1}^{L}{\frac{d^2 -2(m-n)^2 \sigma^2}{\left( d^2 + (m-n)^2 \sigma^2 \right)^{5/2}}}}.
\label{stacked-interaction}
\end{equation}
As shown in Fig.~\ref{stacked-simul}, the interaction is always repulsive, increases at small $d$ as $L$ increases, but seems to become independent of $L$ for larger $d$.

For the staggered geometry, the interaction energy is given by,
\begin{equation}
\frac{U_{g}(d;L)}{ \sigma^3 A_D}=\sum_{n=1}^{L}{\sum_{m=1}^{L}{ \frac{d^2 -2\left(m-n+\frac{1}{2}\right)^2 \sigma^2}{\left( d^2 + \left(m-n+\frac{1}{2}\right)^2 \sigma^2 \right)^{5/2}}}}.
\label{staggered-interaction}
\end{equation}
As shown in Fig.~\ref{staggered-simul}, for small $d$ the staggered interaction is attractive, increasing in an approximately linear way with $L$.  While the range of attraction also grows with $L$, the attractive part is short ranged with a maximum in $U_g(d)$ at $d \approx 1.75\sigma$ for $L=100$.
At larger $d$, the interaction is repulsive and quickly becomes indistinguishable from the stacked arrangement.

One motivation for modeling the system in the way we do is to highlight the short range attraction and long range repulsion between  the basic constituents of the system in the regime where chains are well formed.

\subsection{Energetics of clustering}

The main impact of the short range attraction and long range repulsion is the emergence of energetically favored finite size clusters.  This phenomenon has been the topic of numerous studies on colloidal systems, including studies pertaining to equilibrium thermal gels.

\begin{figure}[ht]
\centering\includegraphics[height=4.8cm, width=7.0cm]{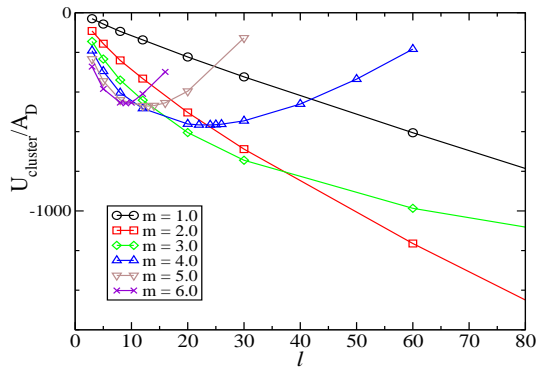}
\caption{(Color online) The potential energy of an isolated cluster of size $m\times l$ as a function of $l$.}
\label{physical-cluster}
\end{figure}

To illustrate the effect for our system, we compose rectangular $bct$ clusters $m$ chains wide and $l$ chains long and plot for each $m$ the potential energy of each cluster as a function of $l$ in Fig.~\ref{physical-cluster}.  All chains in the cluster have length $L=50$ and are in contact with nearest neighbors, i.e. $d/\sigma=\sqrt{3}/2\approx0.866$.  As the figure illustrates, the potential energy for narrow clusters ($m=1$, 2, 3) keeps decreasing with length, while for wider clusters ($m\ge 4$), there appears an energetically favored length.  For square $m\times m$ clusters (not shown) potential energy increases beyond $m=7$.

While this analysis is limited because it neglects interaction with other clusters (i.e., density effects), as well as entropic considerations, it does foreshadow the importance of long, string-like structures as well as more compact clusters of finite size.

\subsection{Monte Carlo simulations}

The system of interacting chains is essentially two dimensional, and as such, we represent each chain of colloidal spheres as a disk in a plane.  To capture the short range structure, we label disks with either a 1 or -1.  If the product of the labels is 1, then the interaction is given by $U_c(d;L)$, the always repulsive stacked potential.  Otherwise, the disks interact through $U_g(d;L)$, which is deeply attractive at short range and repulsive for distances beyond $1.64\sigma$. 
We choose all chains to be composed of $L=50$ particles, and we simulate $N_p=2500$ disks.

We follow the usual Metropolis MC algorithm to generate new configurations at constant area~\cite{frenkel}.  
To model the slight chain motion in the $z$ direction that can change local packing (staggered and stacked), we choose between two MC moves, each with 50\% probability: a random displacement of a disk, or a random displacement of a disk accompanied by a change in label of that disk.
The simulations are done for a broad range of $T^*$ and area fraction $A_\phi \equiv N_p \pi \sigma^2/ (4 {L_B}^2)$, where $L_B$ is the square simulation box length, and we report quantities for state points where the mean square displacement of the disks after an initial equilibration period is at least $\sigma^2$.  We truncate the potentials at $r_c=L_B/2$.  For the largest $A_\phi=70\%$, $r_c = 26.48\sigma$.

To charcterize our system, we calculate the radial distribution function $g(r)$, the structure factor $S(q)=\left< \rho_q \rho_{-q} \right>$, where $\rho_{\vec{q}} = N_p^{-1/2} \sum_{i=1}^N \exp{\left(-i  \vec{q} \cdot \vec{r}_i  \right)}$, and 
the potential energy of the system $U(T,A_\phi)$, which includes the standard tail correction assuming $g(r>r_c)\approx 1$.
We aslo calculate the distribution of  clusters $N(n)$, i.e., the number of clusters of size $n$.  
We do not consider the degree of local crystallinity in defining a cluster, but rather simply define two disks to be in the same cluster if they are within
a distance of  $r_n=1.233\sigma$ of each other [$U_g(r_n)=0$].  
We do so in the spirit of Toledano and coworkers~\cite{toledano}, who studied clustering in a system with competing short range attraction and long range repulsion in three dimensions, and used the appearance of a local maximum in $N(n)$ for $n>0$ to define a cluster phase.  This maximum is equivalently a minimum in the free energy of forming an $n$-sized cluster, $\Delta F(n)=-\ln{[ N(n)/N(1) ]}$.
We also keep track of whether or not there is a system-spanning cluster and define the percolation line by identifying state points that have a 50\% chance of containing a spanning cluster.

We report the pressure in terms of $g(r)$ via,
\begin{eqnarray}
P &=& \rho k_B T + \frac{1}{2} k_B T\pi\rho^{2}\sigma^{2}g_{c}(\sigma)\nonumber \\
  &+& \frac{1}{2} k_B T\pi\rho^{2}(0.866\sigma)^{2}g_{g}(0.866\sigma)\nonumber \\
  &-& \frac{1}{2}\pi\rho^{2}\int_{\sigma}^{r_{c}}r^2 \frac{dU_{c}(r)}{dr} g_c(r)dr \nonumber \\
  &-& \frac{1}{2}\pi\rho^{2}\int_{0.866\sigma}^{r_{c}}r^2 \frac{dU_{g}(r)}{dr} g_g(r)dr \nonumber \\
  &-& \frac{1}{2}\pi\rho^2 \int_{r_{cut}}^{\infty} r^2  \frac{dU_{c}(r)}{dr} dr \nonumber \\
  &-& \frac{1}{2}\pi\rho^2 \int_{r_{cut}}^{\infty} r^2  \frac{dU_{g}(r)}{dr} dr,
\label{pressure}
\end{eqnarray} 
where $\rho$ is the number density per unit area, $g_c(r)$ and $g_g(r)$ are the partial radial distribution functions for stacked and staggered interactions, respectively.  We have accounted for hard core repulsion, omitted the (density dependent) impulsive correction at $r_c$, and have added the long range tail correction.  In this way we approximate the behavior of the untruncated system.

As mentioned in the Introduction, Ref.~\cite{dijkstra} indicates that at low $T$ and over a range in density, the crystal phase 
coexists with a rare fluid.  As another check for coexistence within our system in  addition to tracking $P$, we wish to track the reduced number density $\rho_f$ of the fluid (non-crystal-like) portion of the system as $A_\phi$ varies.  

To determine the number $N_c$ of $bct$-like particles, i.e., disks in a locally crystalline environment, we employ the method developed by Frenkel and co-workers \cite{frenkel1996}, 
based on the local bond-order analysis that was introduced in Ref.\cite{steinhardt}. 
For each particle we define the quantity 
\begin{equation}
q_{lm}(i) = \frac{1}{N_{b}(i)} \sum_{j=1}^{N_{b}(i)} Y_{lm}(\hat{r}_{ij}),
\label{complex-vectors}
\end{equation}
where the sum is carried out over all neighbouring particles $N_{b}(i)$ that are within a distance $1.05\sigma$, 
$Y_{lm}(\hat{r}_{ij})$ are the spherical harmonics calculated for the normalized direction vector $\hat{r}_{ij}$ between the neighbours.
The unit vector $\hat{r}_{ij}$ is determined by the polar and azimuthal angels $\theta_{ij}$ and $\phi_{ij}$, although here in
two dimensions $\theta_{ij}\equiv \pi/2$.  Since our crystallites posses square symmetry, we use $l=4$.
For each pair of neighboring particles $i$ and $j$, we calculate the correlation $\sum_{m=-4}^{4} \hat{q}_{4m}(i) \hat{q}^{*}_{4m}(j)$, 
where $\hat{q}_{4m} = q_{4m}(i) / [ \sum_{m=-4}^{4}|q_{4m}(i)|^{2} ]^{1/2}$ and $q^*$ is the complex conjugate of $q$. 
If the correlation between two neighbouring particles is greater than $0.05$ or less than $-0.05$, then particles $i$ and  $j$ are considered to be {\it connected}. 
If a particle has at least $3$ connected neighbours, then it is considered a crystal-like particle.

Additionally, since we observe crystallites connected through a single bridging particle,
we also define a particle to be crystal-like if it has only two neighbors that are both {\it connected} to it and the angle between the separation vectors connecting the particle with the two neighbours is less than $\alpha$.  The choice of $\alpha$ is somewhat arbitrary, and so we report results for 
$\alpha = 103^\circ$ and $\alpha = 115^\circ$.  This procedure labels as crystal-like corner particles of crystallites as well as compact squares of four particles.

To determine $\rho_f$, we first define an area occupied by crystallites $A_c$ by associating a square of length $\sigma$ to each crystal-like disk and accounting for overlaps.  The area occupied by the fluid is $A_f=L_B^2 - A_c$ and  $\rho_f= (N_p - N_c) \sigma^2/A_f$.  $\rho_f$ should be constant with respect to $A_\phi$ along an isotherm in a region of crystal-fluid coexistence.

Finally, to test for the existence of a hexatic phase, we report on the orientational correlation function
\begin{eqnarray}
G_4(r) & = & \left< q_4(r) q^*_4(0) \right>, \\
q_4(\vec{r}_j) & = & \frac{1}{N_b(j)} \sum_{k=1}^{N_b(j)} \exp{\left( 4 i \theta_{jk} \right)}
\end{eqnarray}
where $\theta_{jk}$ is the angle formed by the bond between particle $j$ and  neighbor $k$ and an arbitrary fixed axis.
In a crystal, $G_4(r)$ tends to a constant as $r$ increases. In a hexatic phase, it decays as a power law while in a liquid it decays exponentially.

\section{Results}

\begin{figure}[htp]
\centering
\subfigure{
\includegraphics[clip=true, trim=0 0 0 0, height=2.6cm, width=2.6cm]{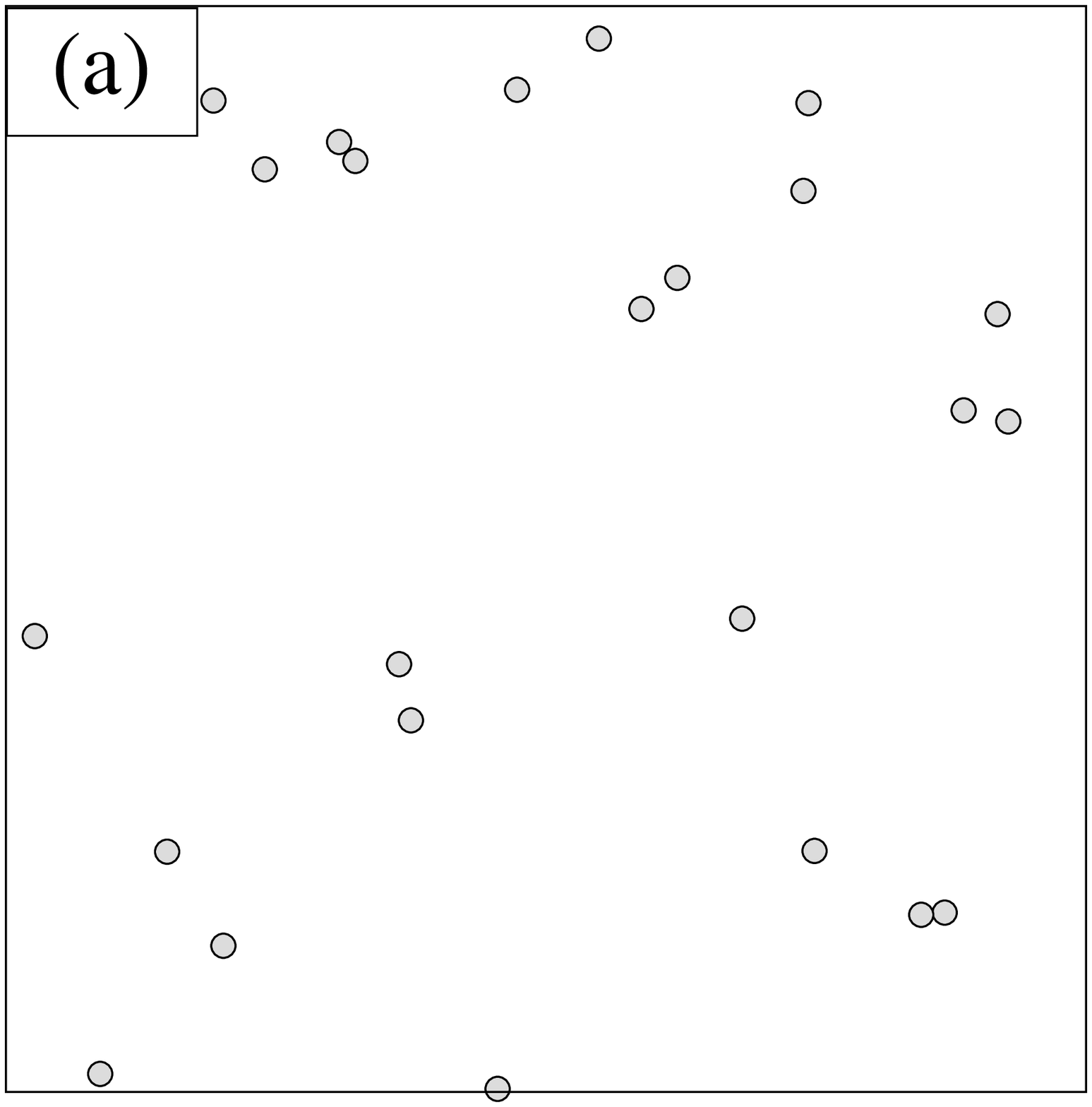}
\label{isochoric-01-temp-40}
}
\subfigure{
\includegraphics[clip=true, trim=0 0 0 0, height=2.6cm, width=2.6cm]{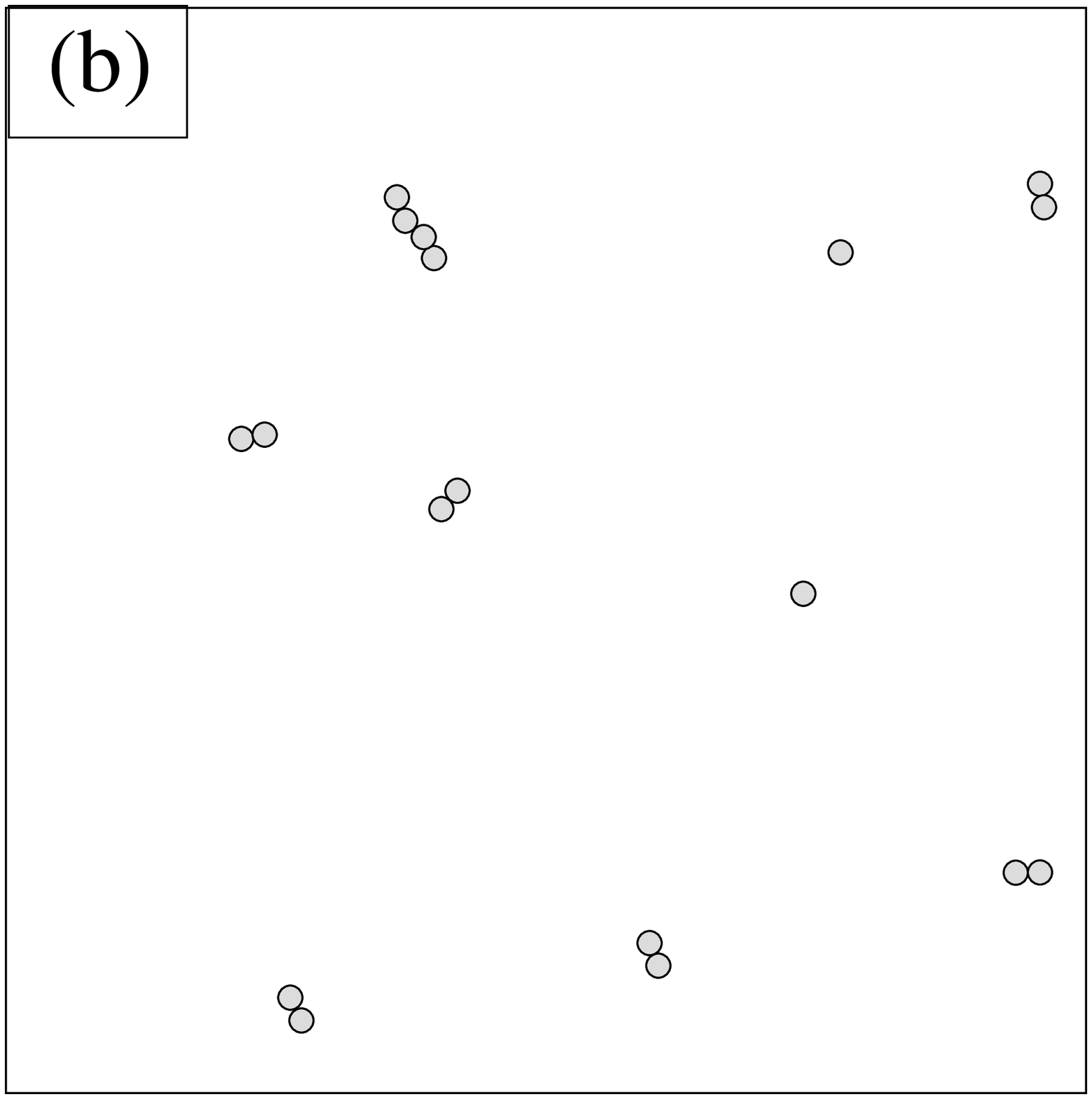}
\label{isochoric-01-temp-18}
}
\subfigure{
\includegraphics[clip=true, trim=0 0 0 0, height=2.6cm, width=2.6cm]{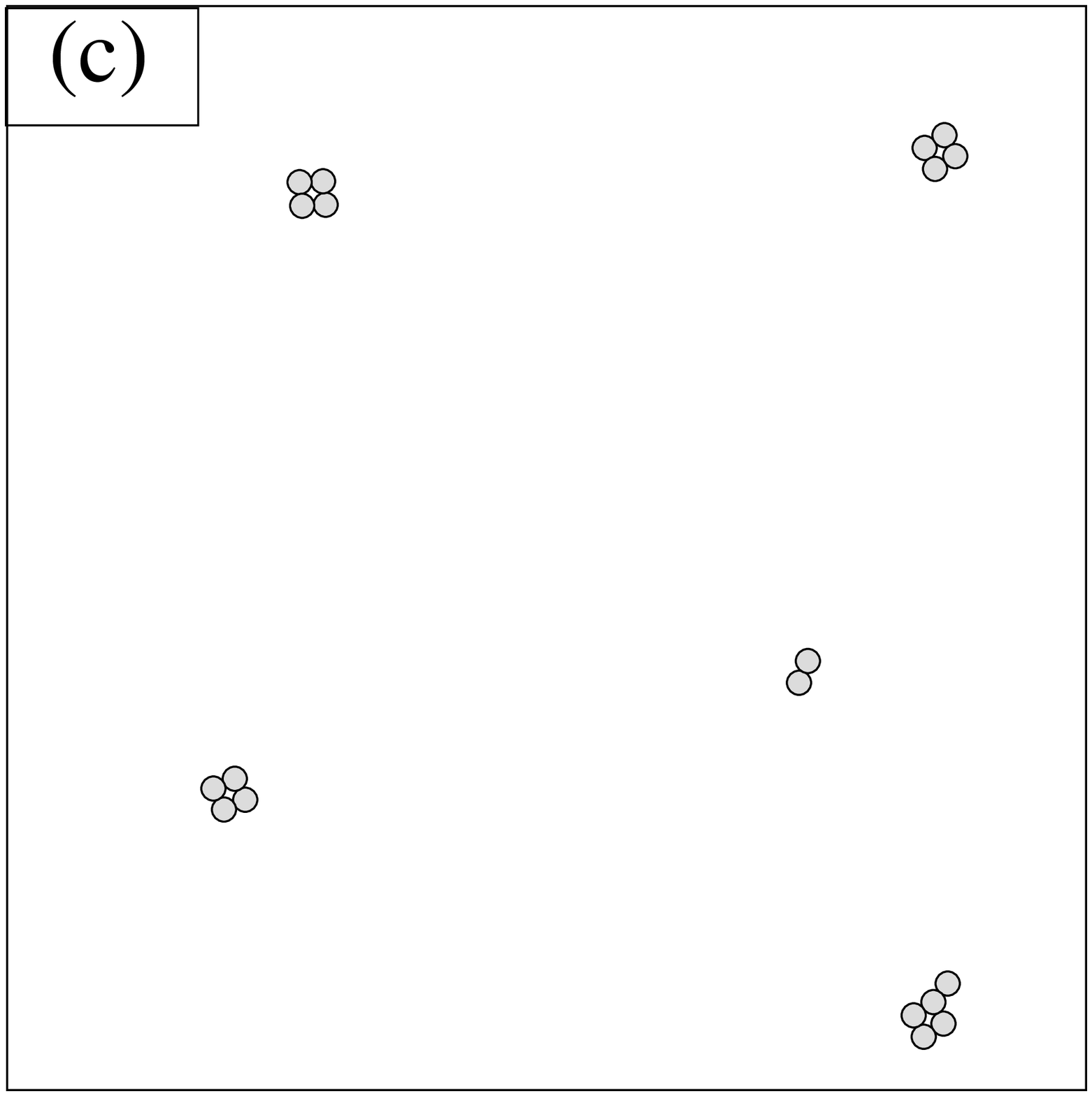}
\label{isochoric-01-temp-04}
}
\subfigure{
\includegraphics[clip=true, trim=0 0 0 0, height=2.6cm, width=2.6cm]{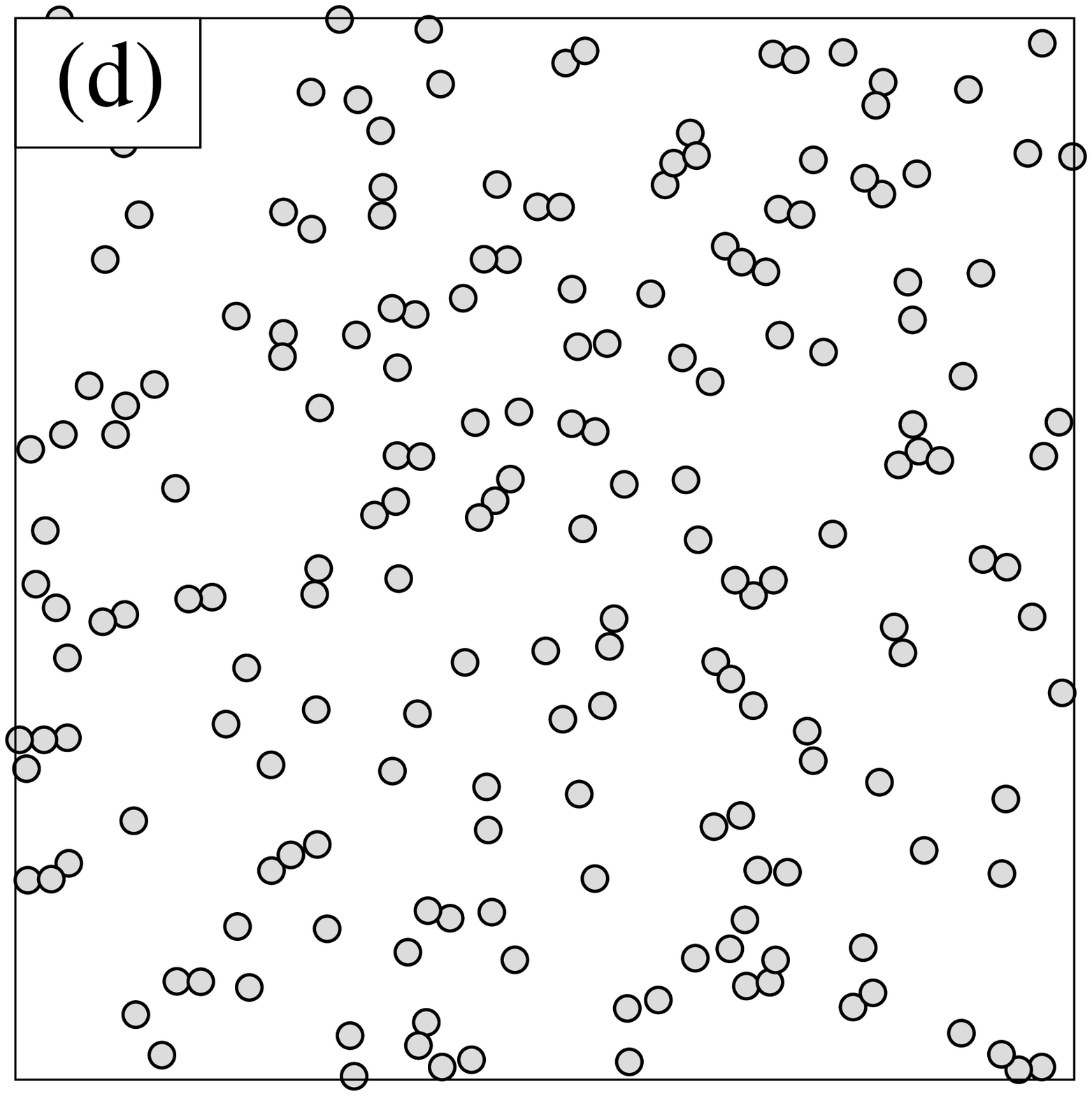}
\label{isochoric-10-temp-40}
}
\subfigure{
\includegraphics[clip=true, trim=0 0 0 0, height=2.6cm, width=2.6cm]{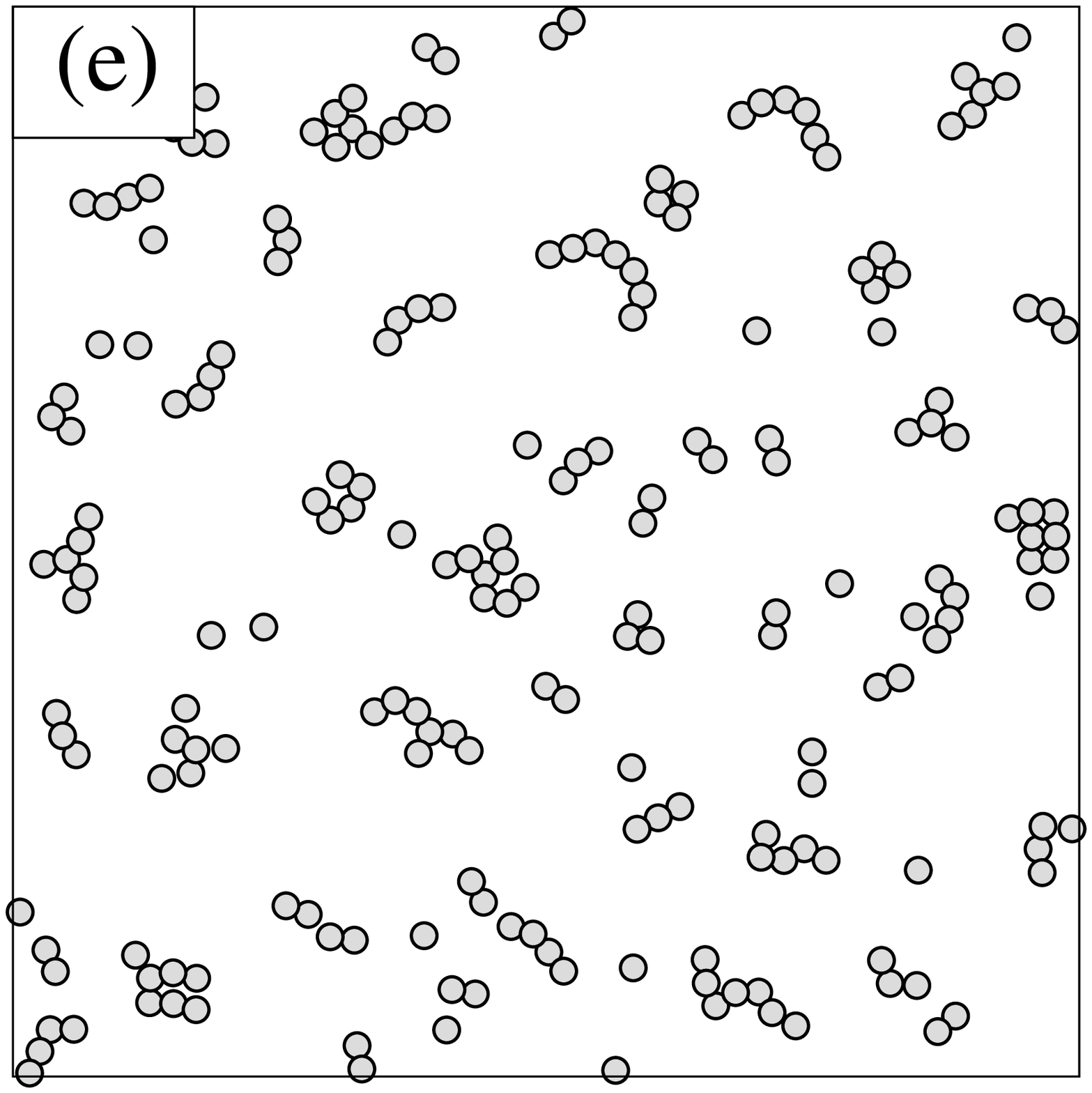}
\label{isochoric-10-temp-23}
}
\subfigure{
\includegraphics[clip=true, trim=0 0 0 0, height=2.6cm, width=2.6cm]{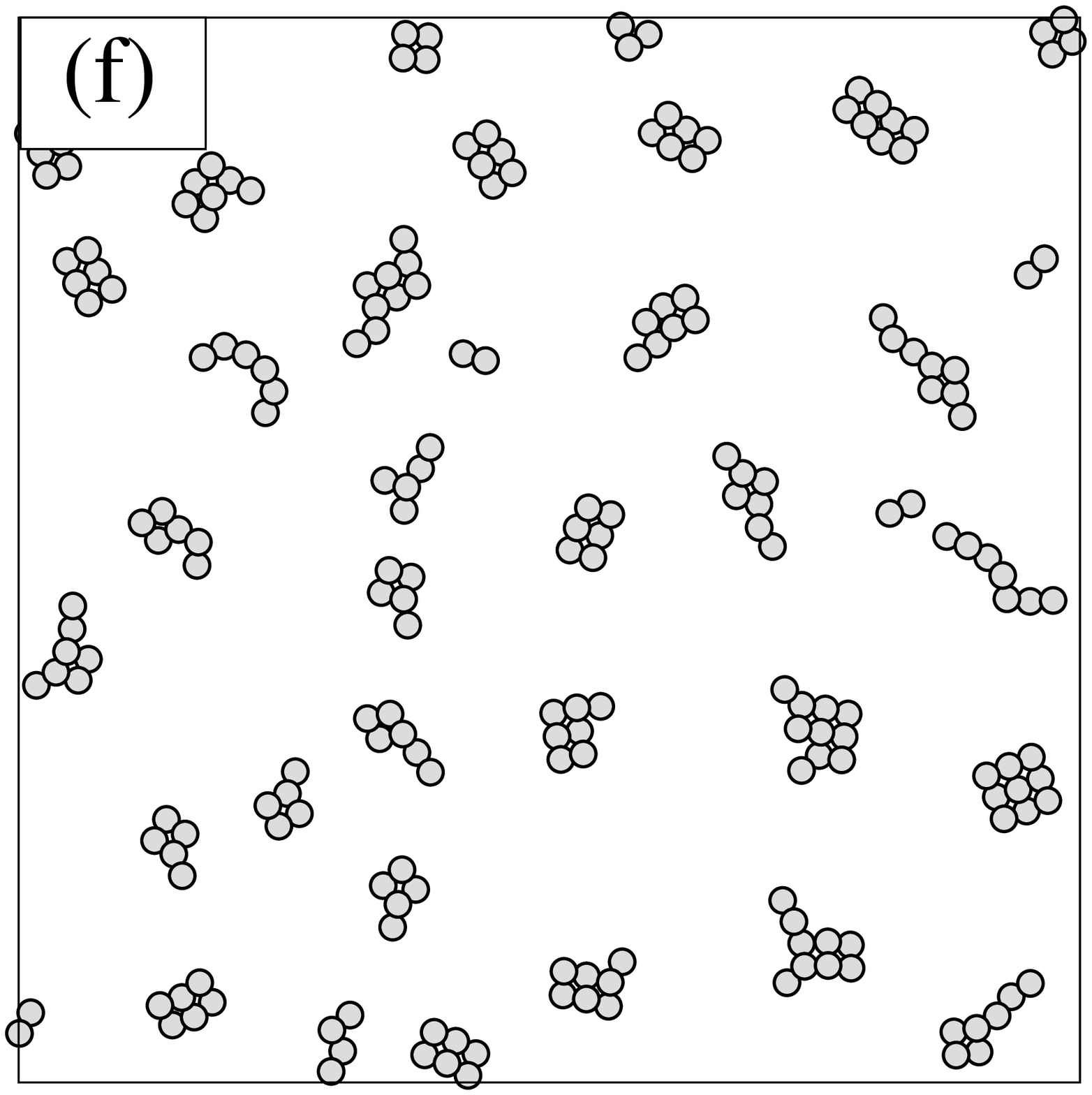}
\label{isochoric-10-temp-15}
}
\subfigure{
\includegraphics[clip=true, trim=0 0 0 0, height=2.6cm, width=2.6cm]{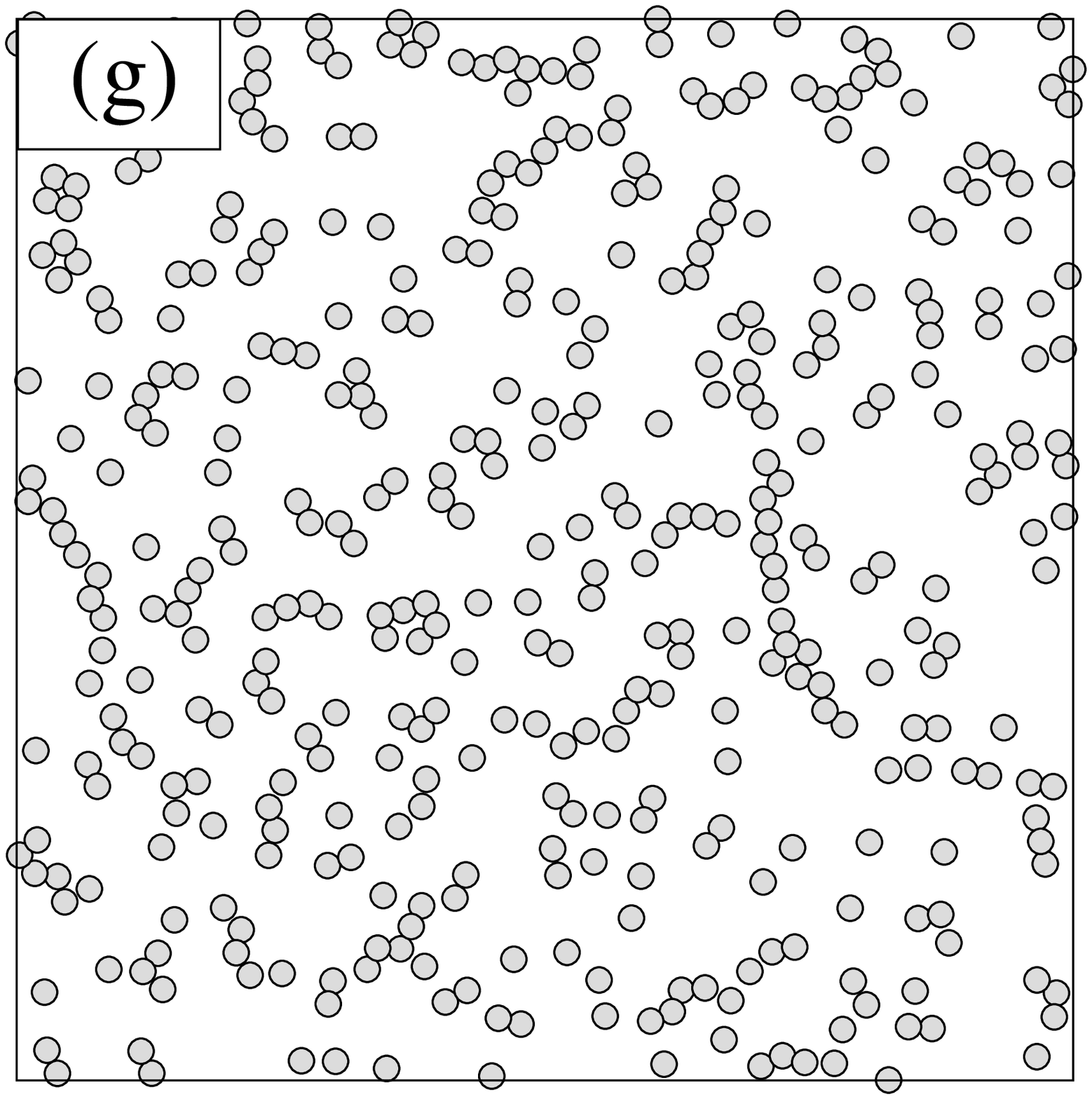}
\label{isochoric-20-temp-40}
}
\subfigure{
\includegraphics[clip=true, trim=0 0 0 0, height=2.6cm, width=2.6cm]{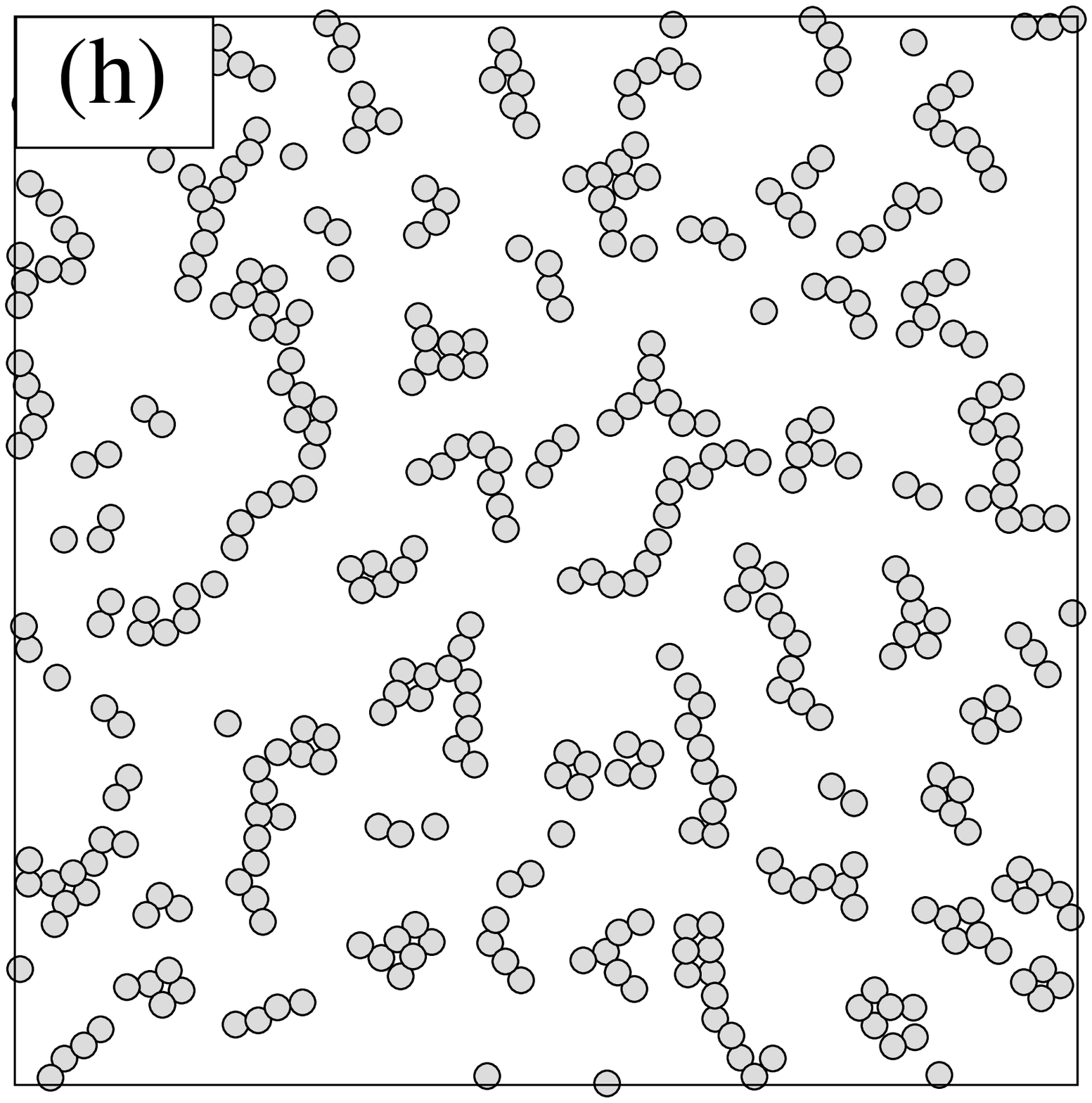}
\label{isochoric-20-temp-23}
}
\subfigure{
\includegraphics[clip=true, trim=0 0 0 0, height=2.6cm, width=2.6cm]{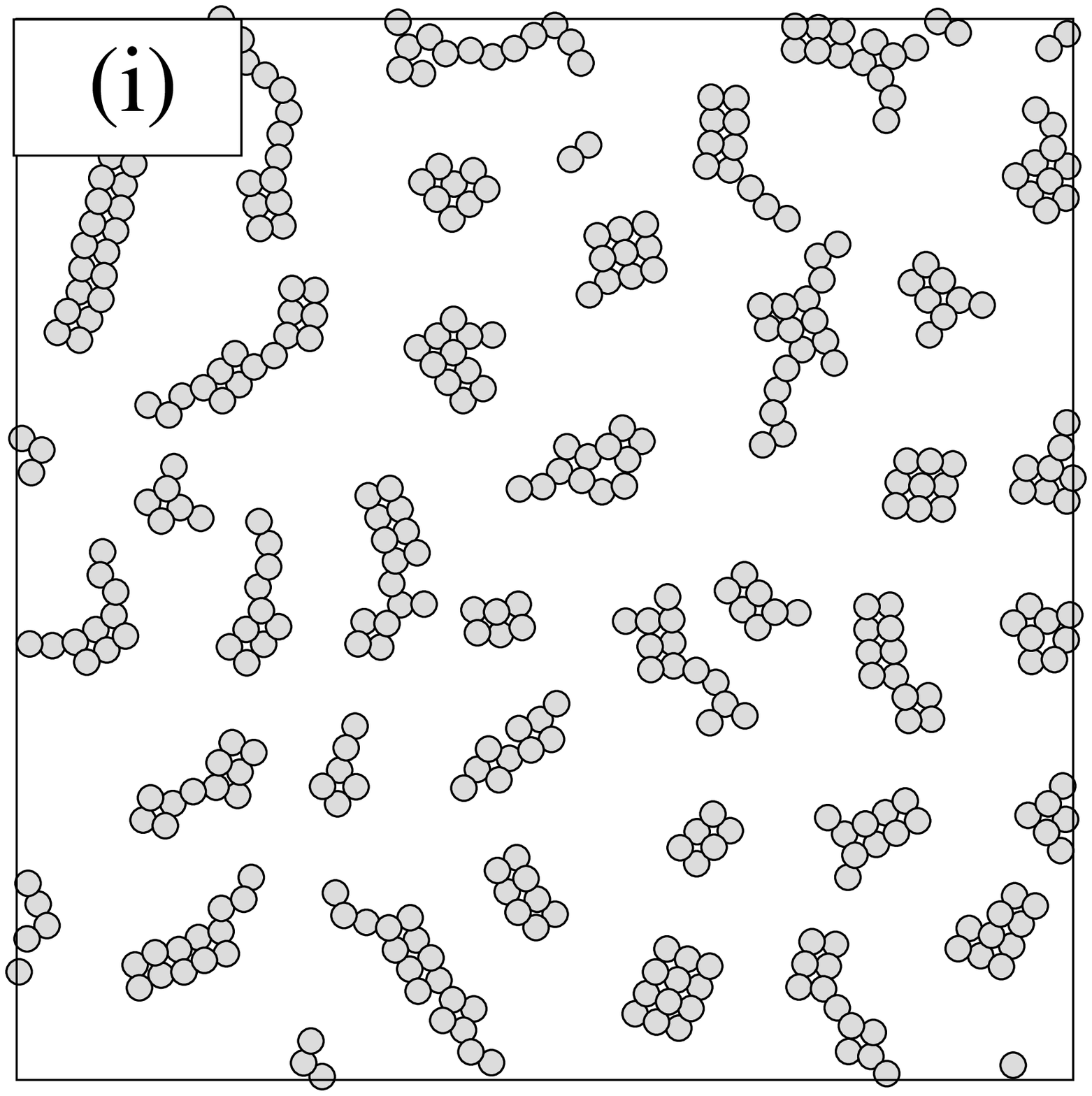}
\label{isochoric-20-temp-15}
}
\subfigure{
\includegraphics[clip=true, trim=0 0 0 0, height=2.6cm, width=2.6cm]{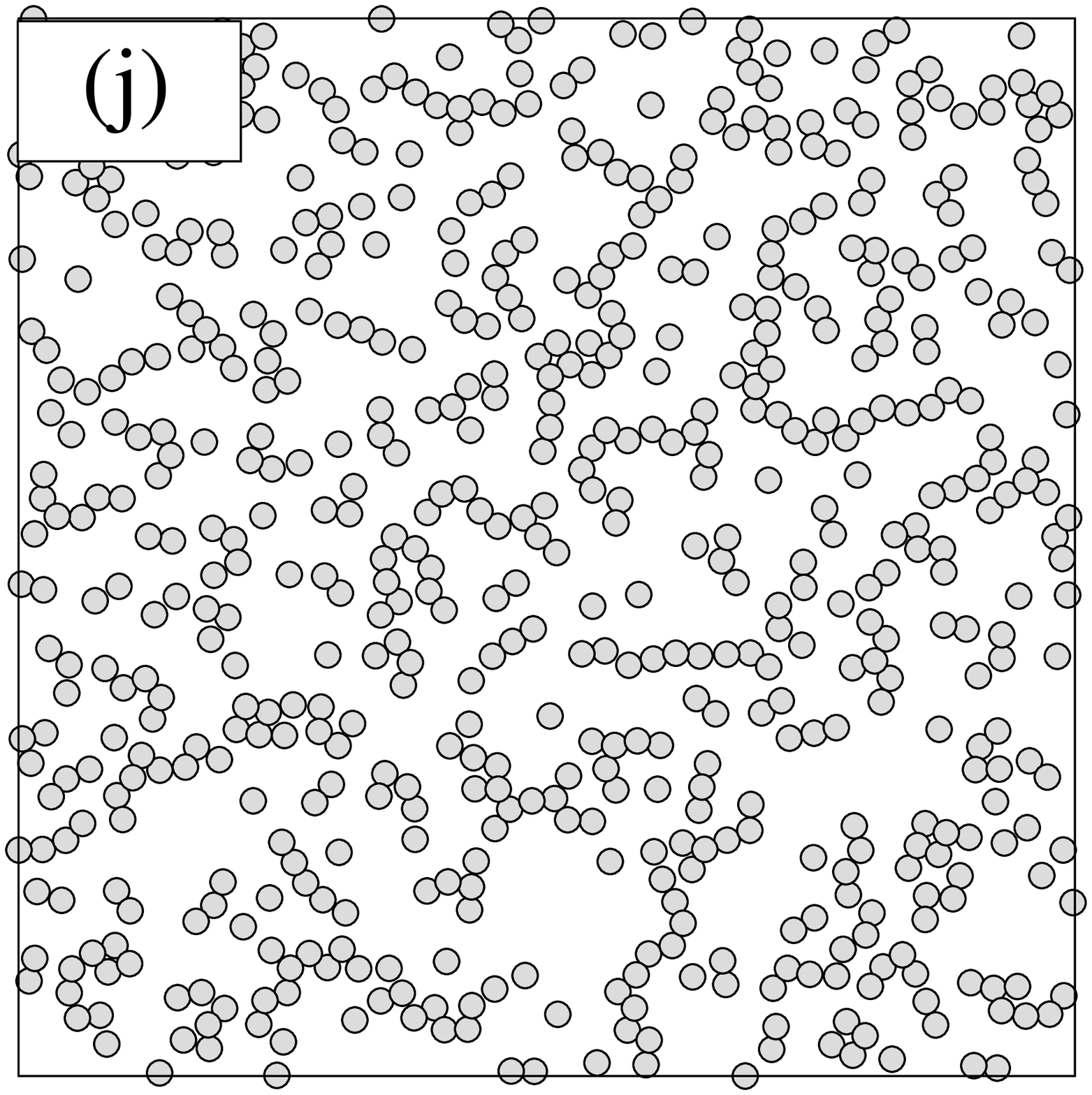}
\label{isochoric-30-temp-40}
}
\subfigure{
\includegraphics[clip=true, trim=0 0 0 0, height=2.6cm, width=2.6cm]{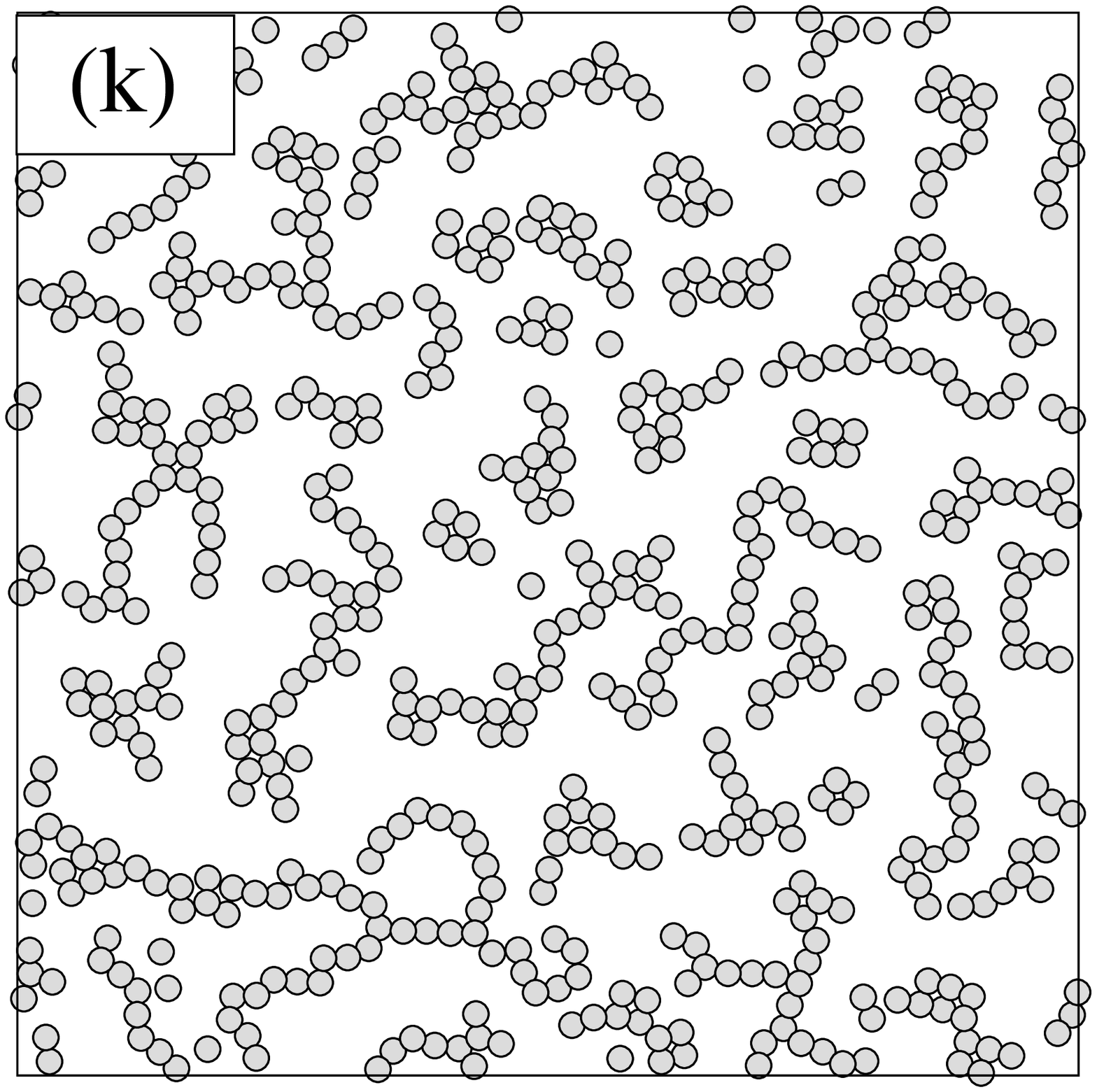}
\label{isochoric-30-temp-23}
}
\subfigure{
\includegraphics[clip=true, trim=0 0 0 0, height=2.6cm, width=2.6cm]{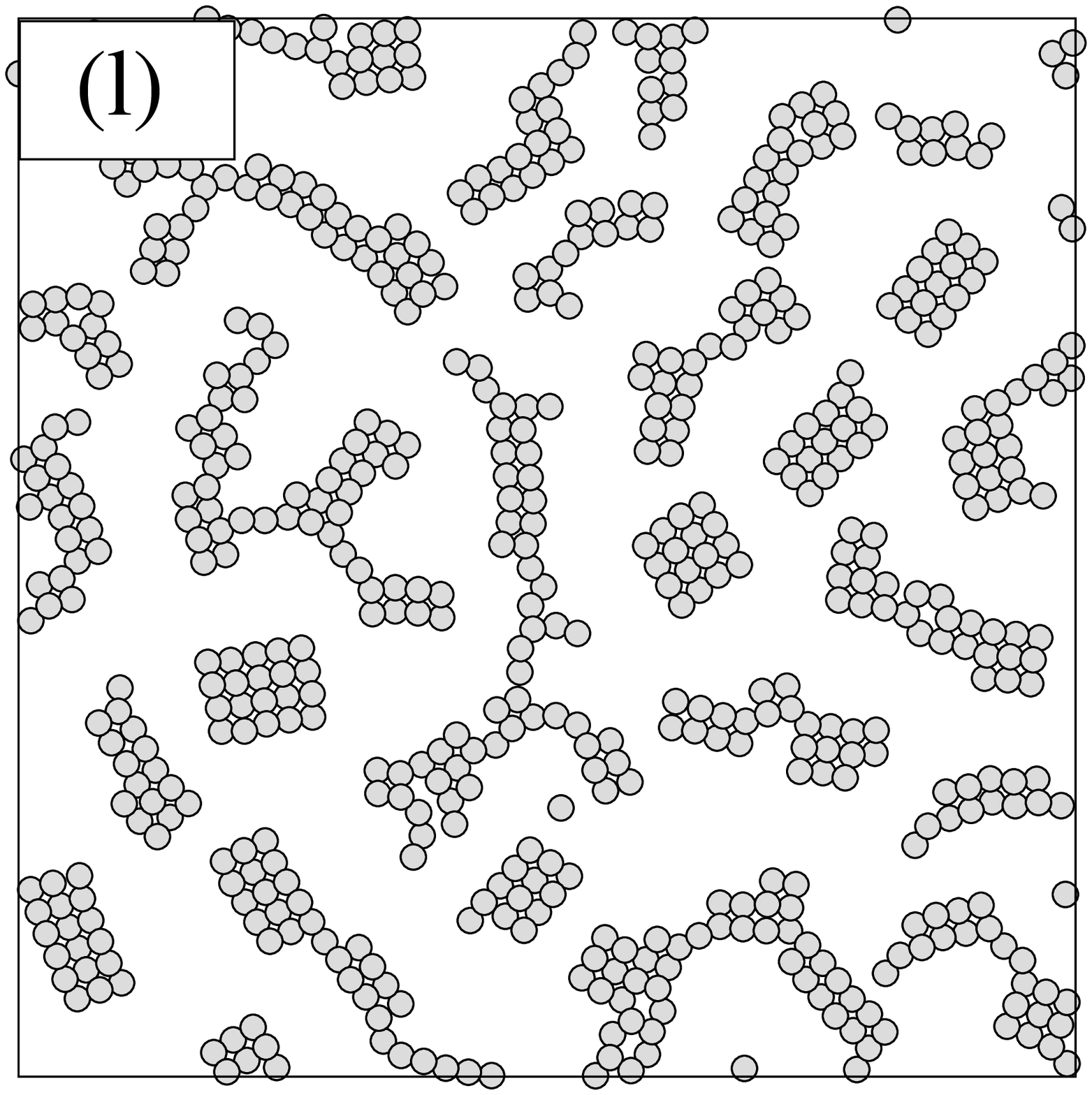}
\label{isochoric-30-temp-15}
}
\caption{Sample configurations for $A_{\phi}$ equal to (a-c) $1\%$, (d-f) $10\%$, (g-i) $20\%$, and (j-l) $30\%$. For $A_{\phi}=1\%$, configurations are shown for $T^*$ equal to \subref{isochoric-01-temp-40} 4.0, \subref{isochoric-01-temp-18} 1.8, and \subref{isochoric-01-temp-04} 0.4. For $A_{\phi}=10\%$, $20\%$, and $30\%$, configurations are shown for $T^*$ equal to  \subref{isochoric-10-temp-40}\subref{isochoric-20-temp-40}\subref{isochoric-30-temp-40} 4.0, \subref{isochoric-10-temp-23}\subref{isochoric-20-temp-23}\subref{isochoric-30-temp-23} 2.3, and \subref{isochoric-10-temp-15}\subref{isochoric-20-temp-15}\subref{isochoric-30-temp-15} 1.5.}
\label{isochore-10}
\end{figure}

\begin{figure}[htp]
\centering
\subfigure{
\includegraphics[clip=true, trim=0 0 0 0, height=2.6cm, width=2.6cm]{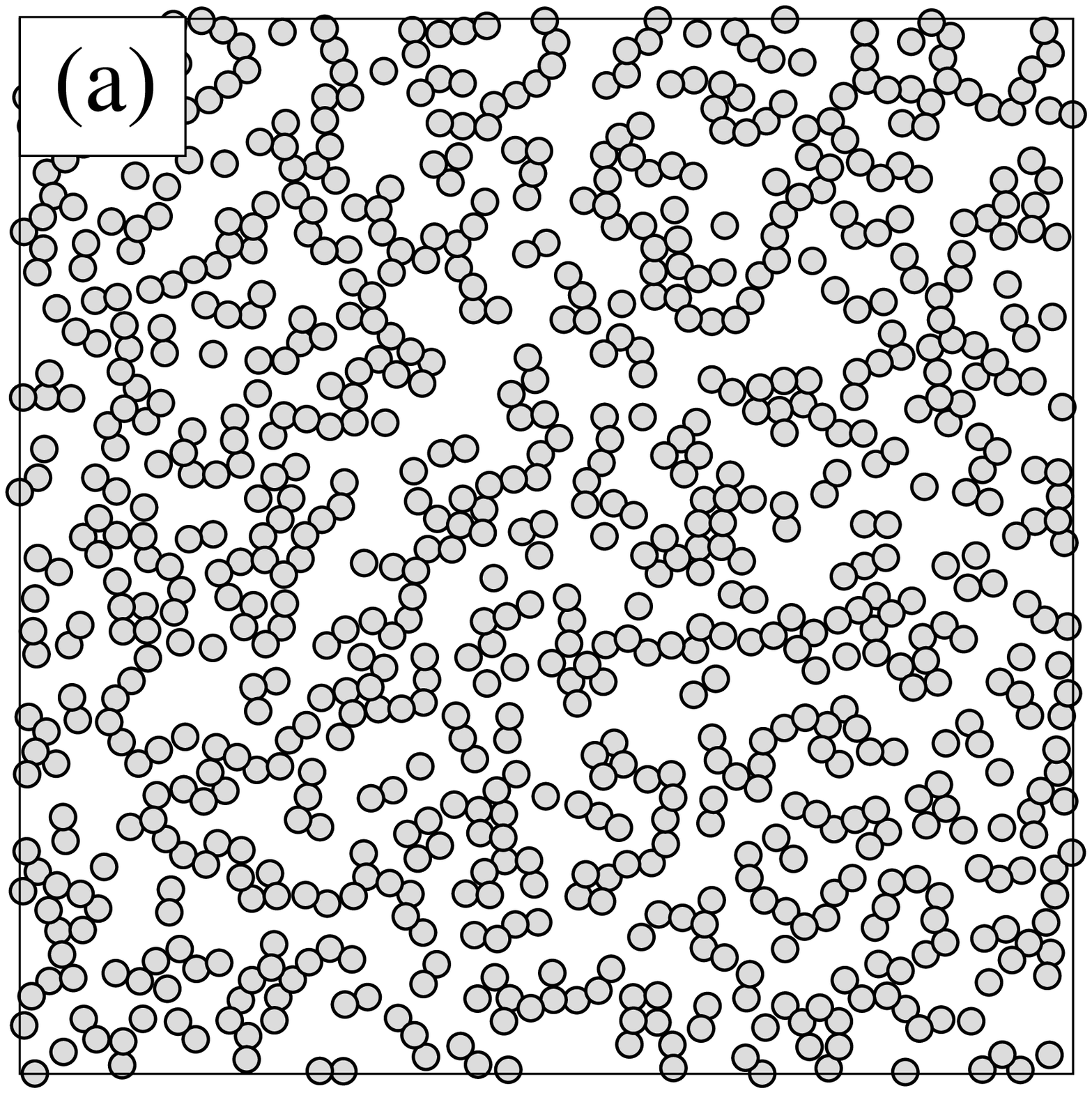}
\label{isochoric-40-temp-40}
}
\subfigure{
\includegraphics[clip=true, trim=0 0 0 0, height=2.6cm, width=2.6cm]{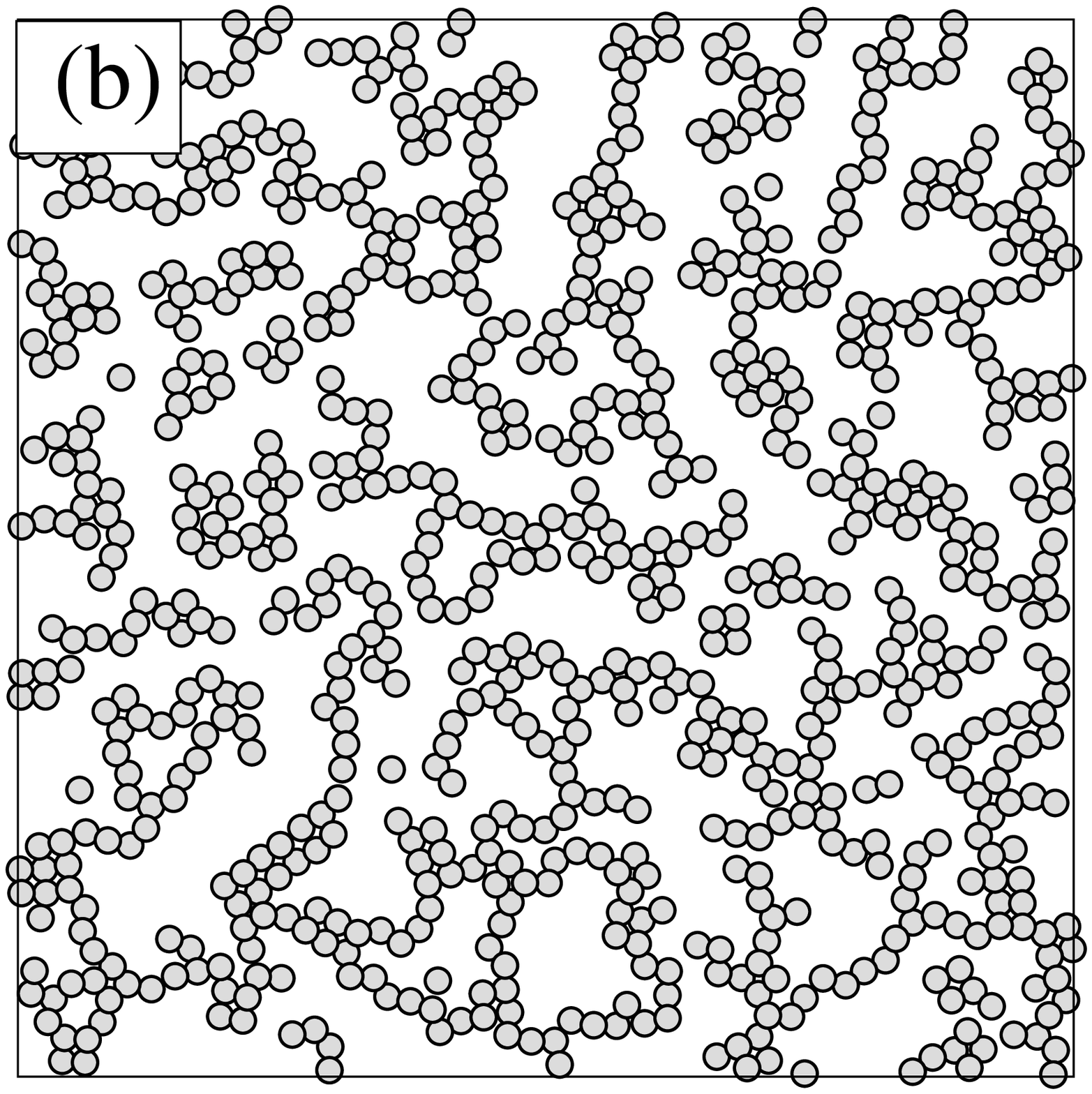}
\label{isochoric-40-temp-23}
}
\subfigure{
\includegraphics[clip=true, trim=0 0 0 0, height=2.6cm, width=2.6cm]{fig6c}
\label{isochoric-40-temp-18}
}
\subfigure{
\includegraphics[clip=true, trim=0 0 0 0, height=2.6cm, width=2.6cm]{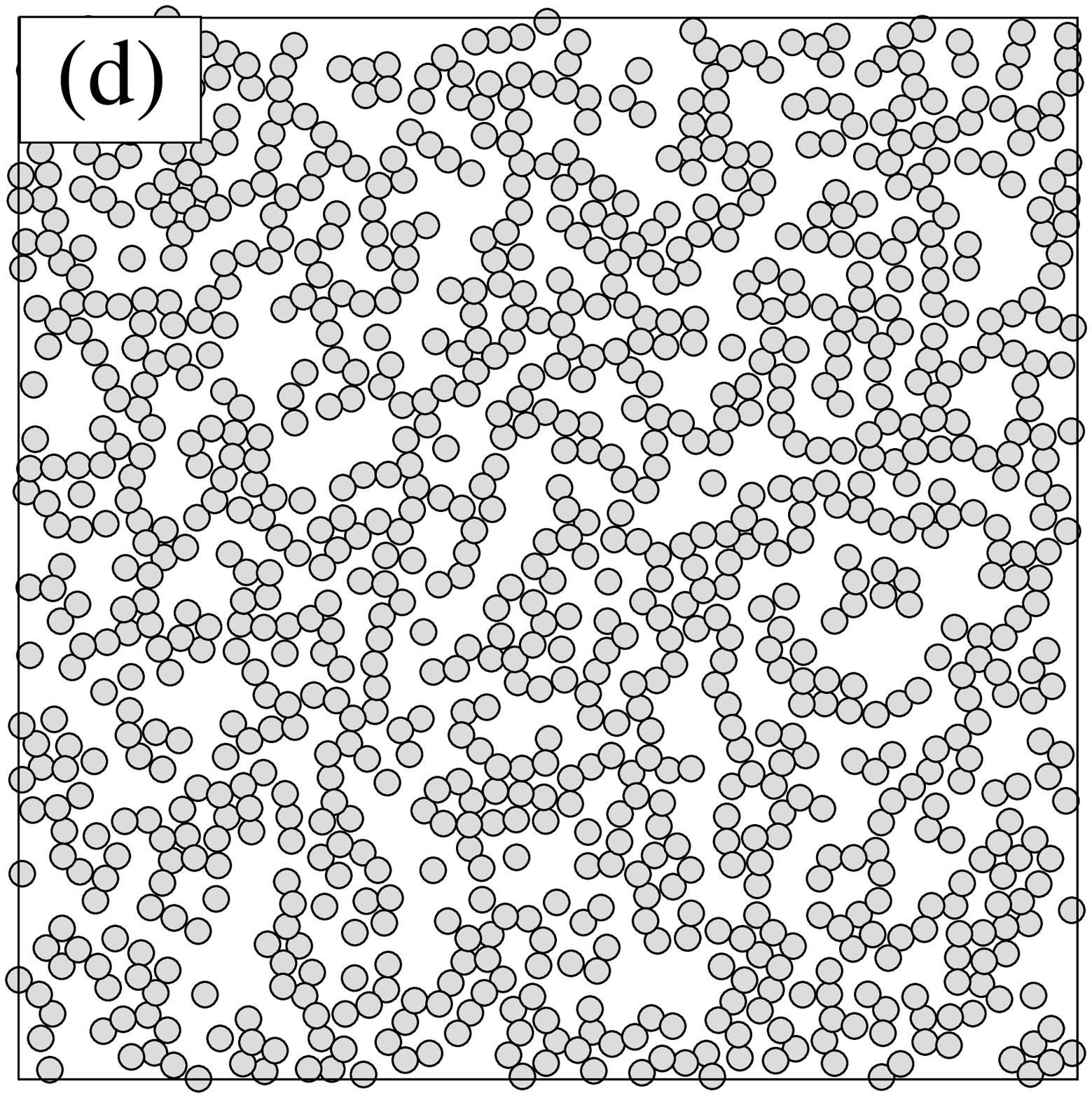}
\label{isochoric-50-temp-40}
}
\subfigure{
\includegraphics[clip=true, trim=0 0 0 0, height=2.6cm, width=2.6cm]{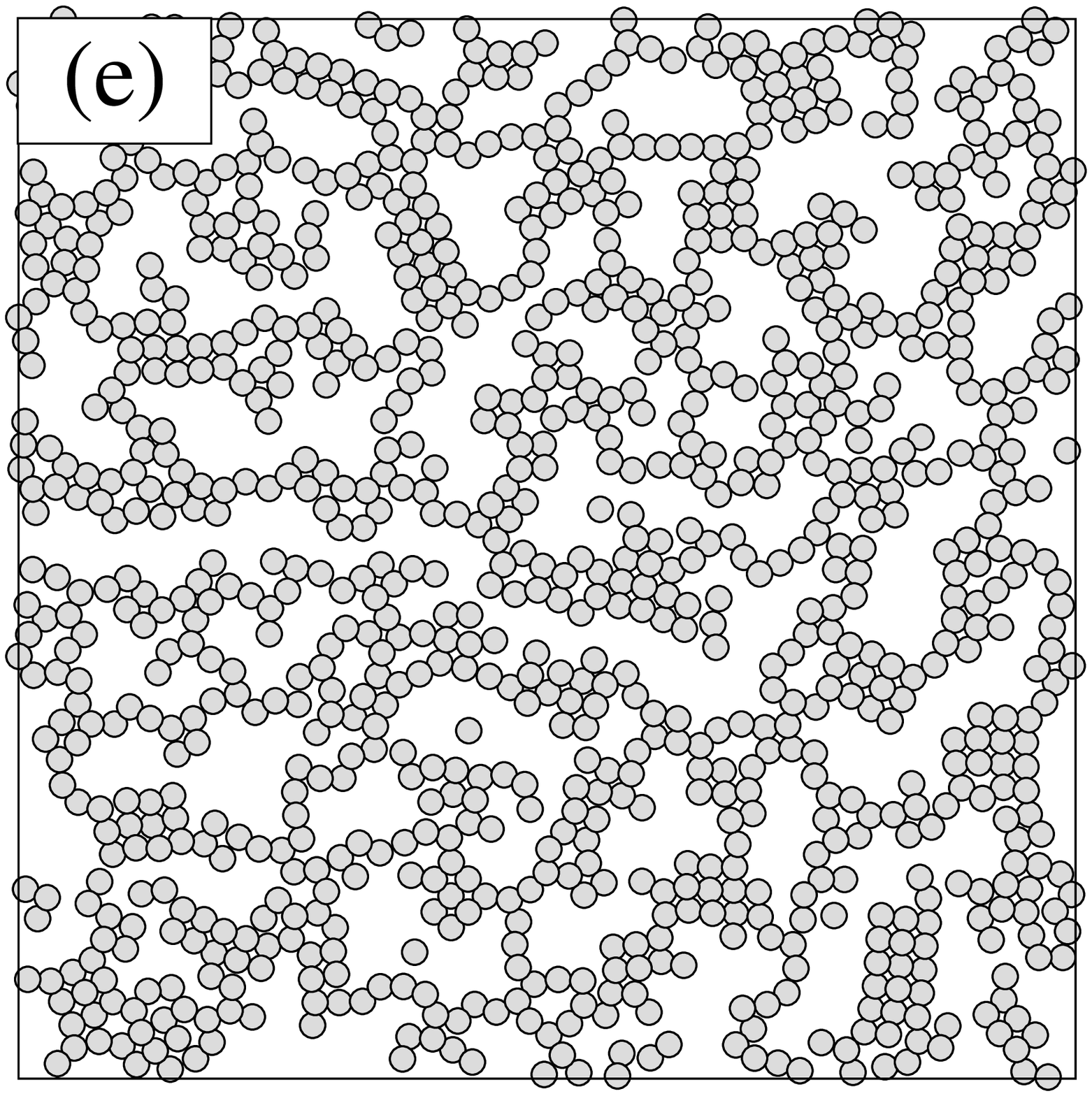}
\label{isochoric-50-temp-23}
}
\subfigure{
\includegraphics[clip=true, trim=0 0 0 0, height=2.6cm, width=2.6cm]{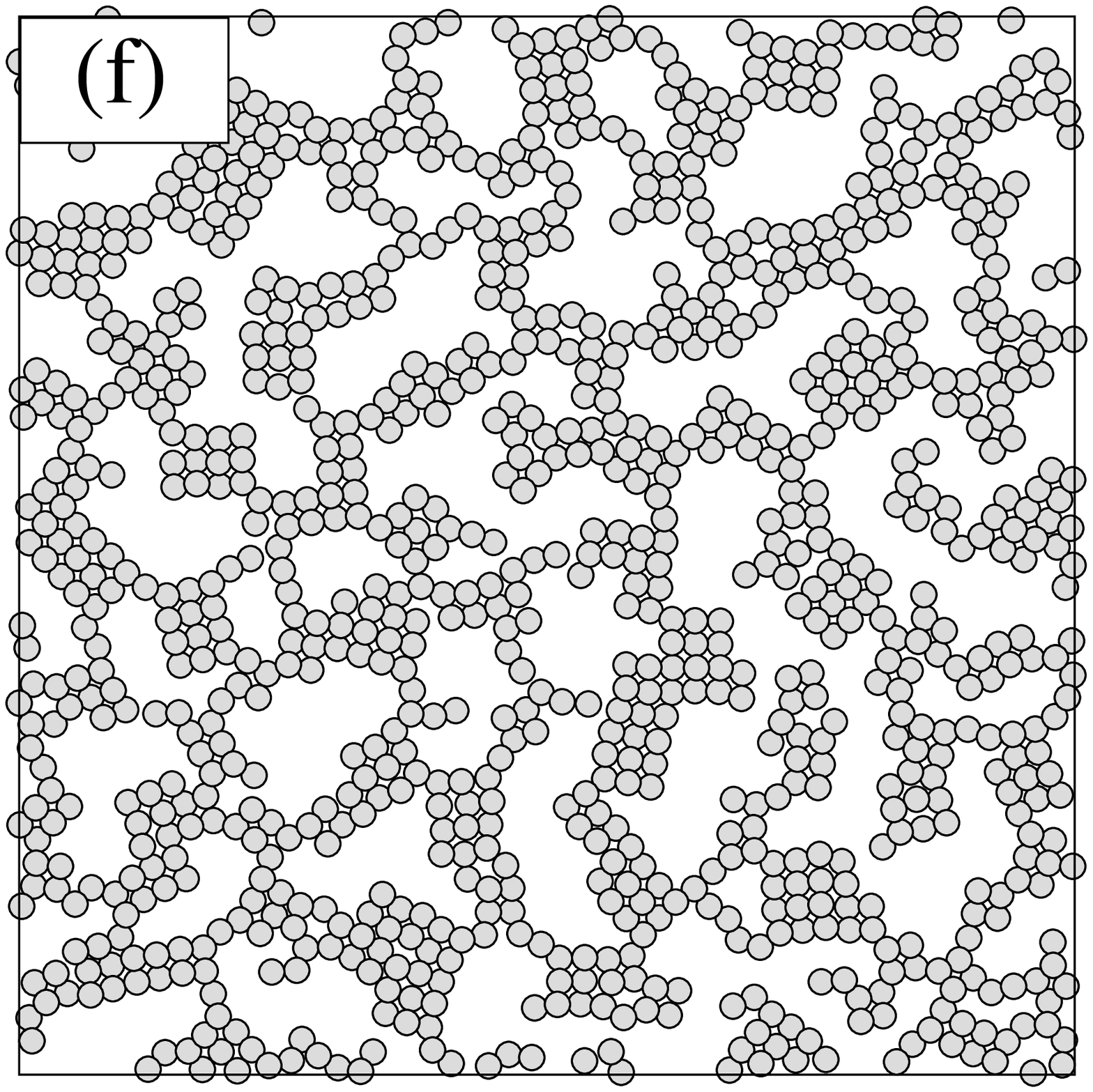}
\label{isochoric-50-temp-18}
}
\subfigure{
\includegraphics[clip=true, trim=0 0 0 0, height=2.6cm, width=2.6cm]{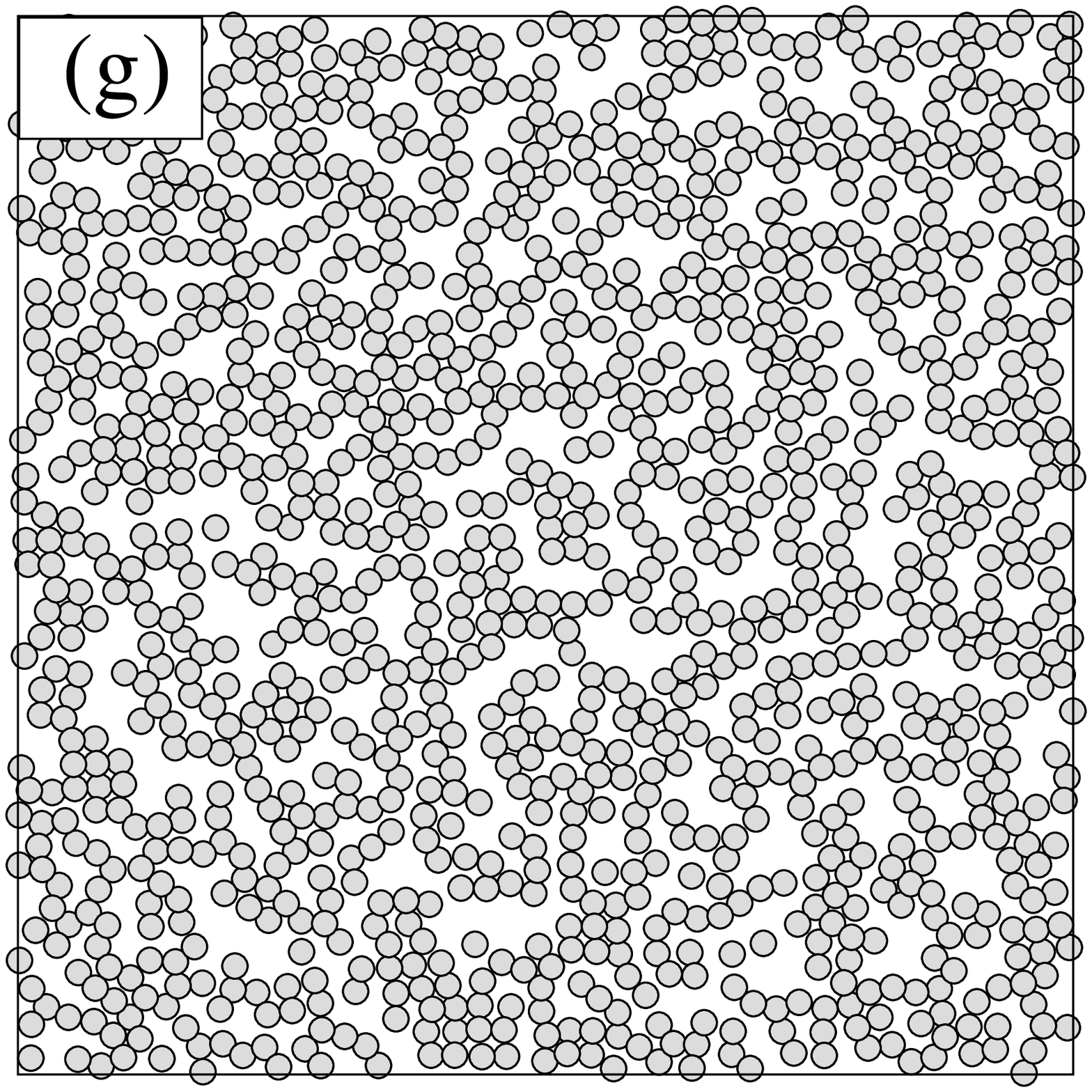}
\label{isochoric-60-temp-40}
}
\subfigure{
\includegraphics[clip=true, trim=0 0 0 0, height=2.6cm, width=2.6cm]{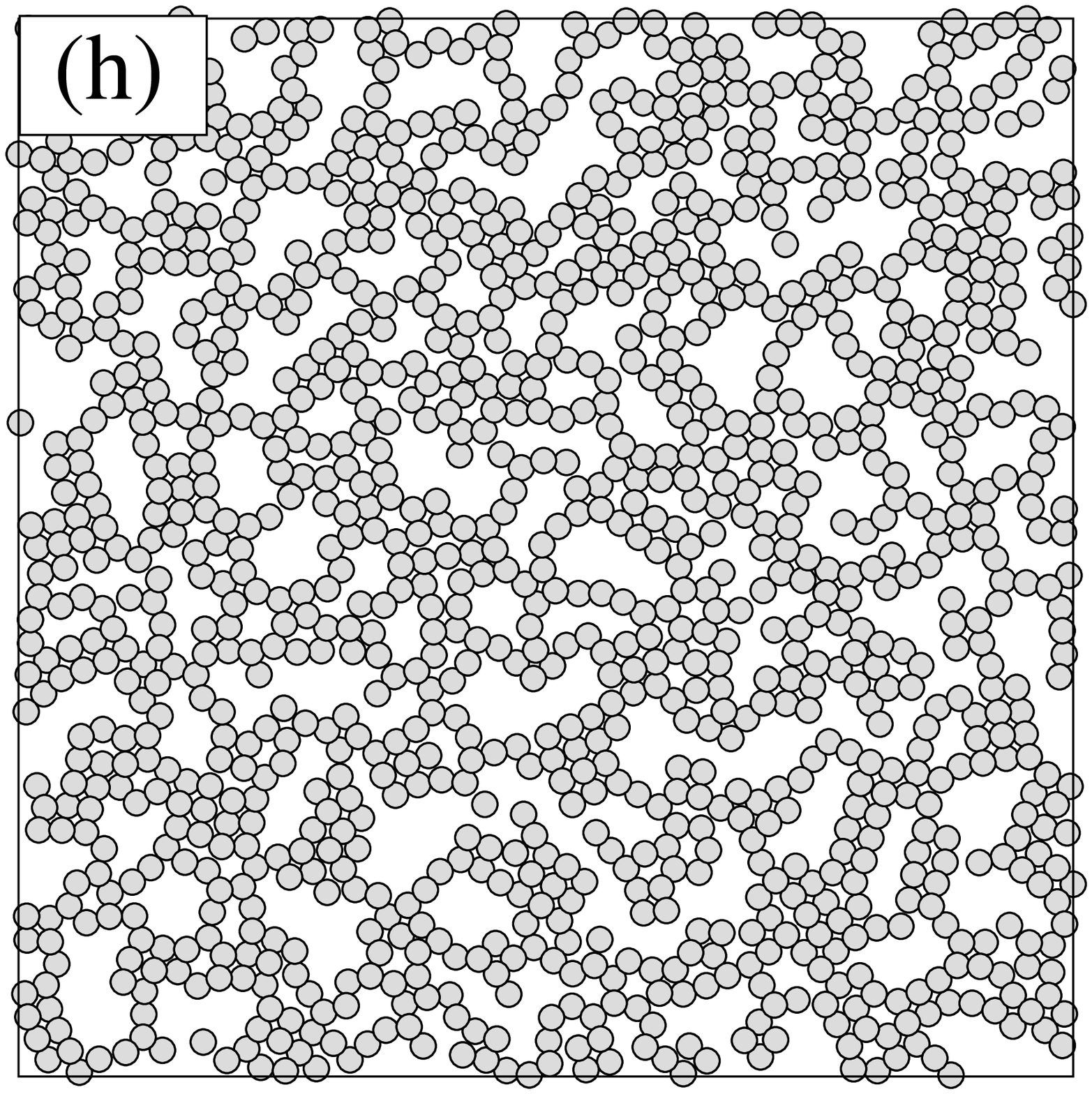}
\label{isochoric-60-temp-23}
}
\subfigure{
\includegraphics[clip=true, trim=0 0 0 0, height=2.6cm, width=2.6cm]{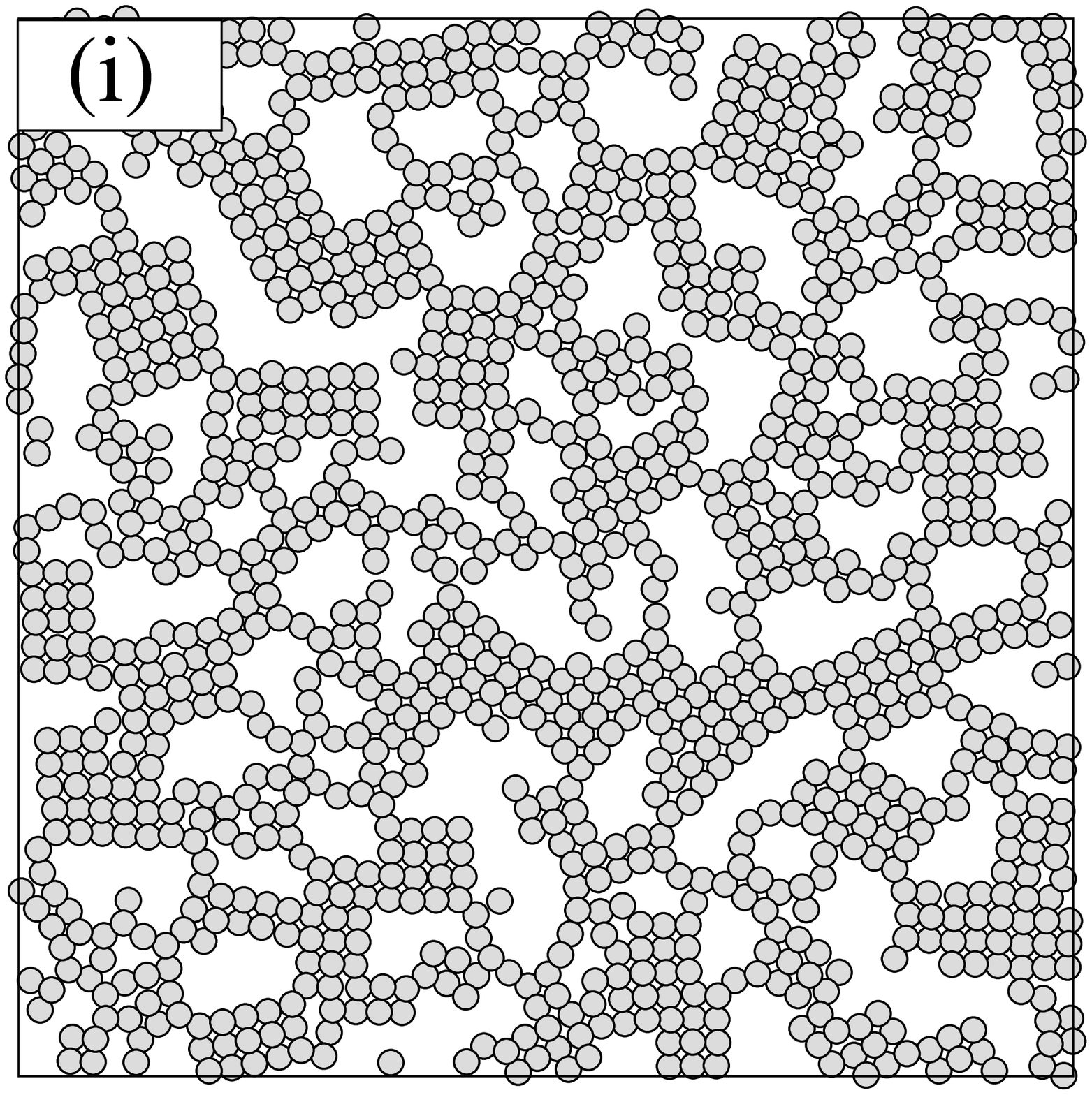}
\label{isochoric-60-temp-18}
}
\subfigure{
\includegraphics[clip=true, trim=0 0 0 0, height=2.6cm, width=2.6cm]{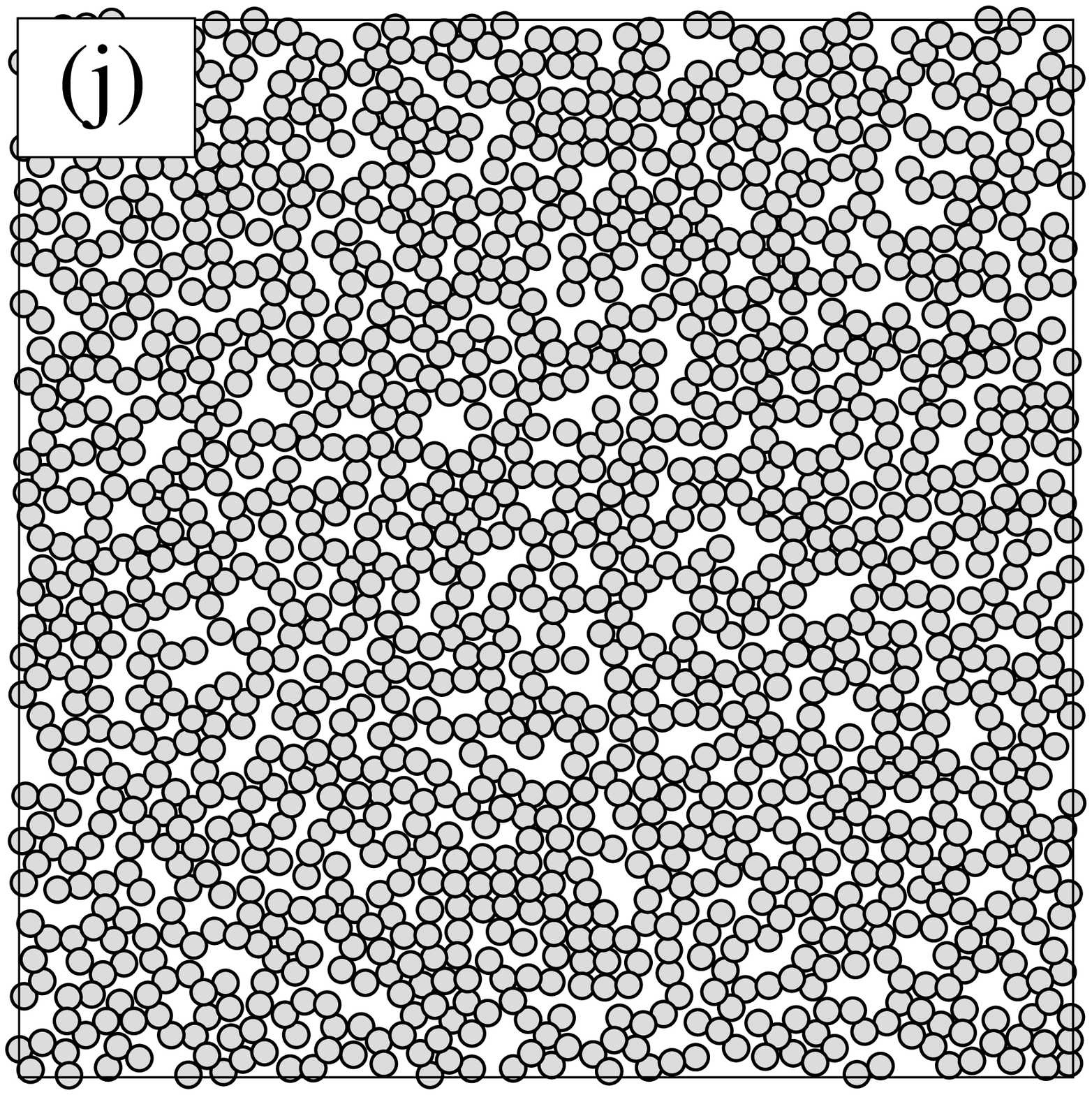}
\label{isochoric-70-temp-40}
}
\subfigure{
\includegraphics[clip=true, trim=0 0 0 0, height=2.6cm, width=2.6cm]{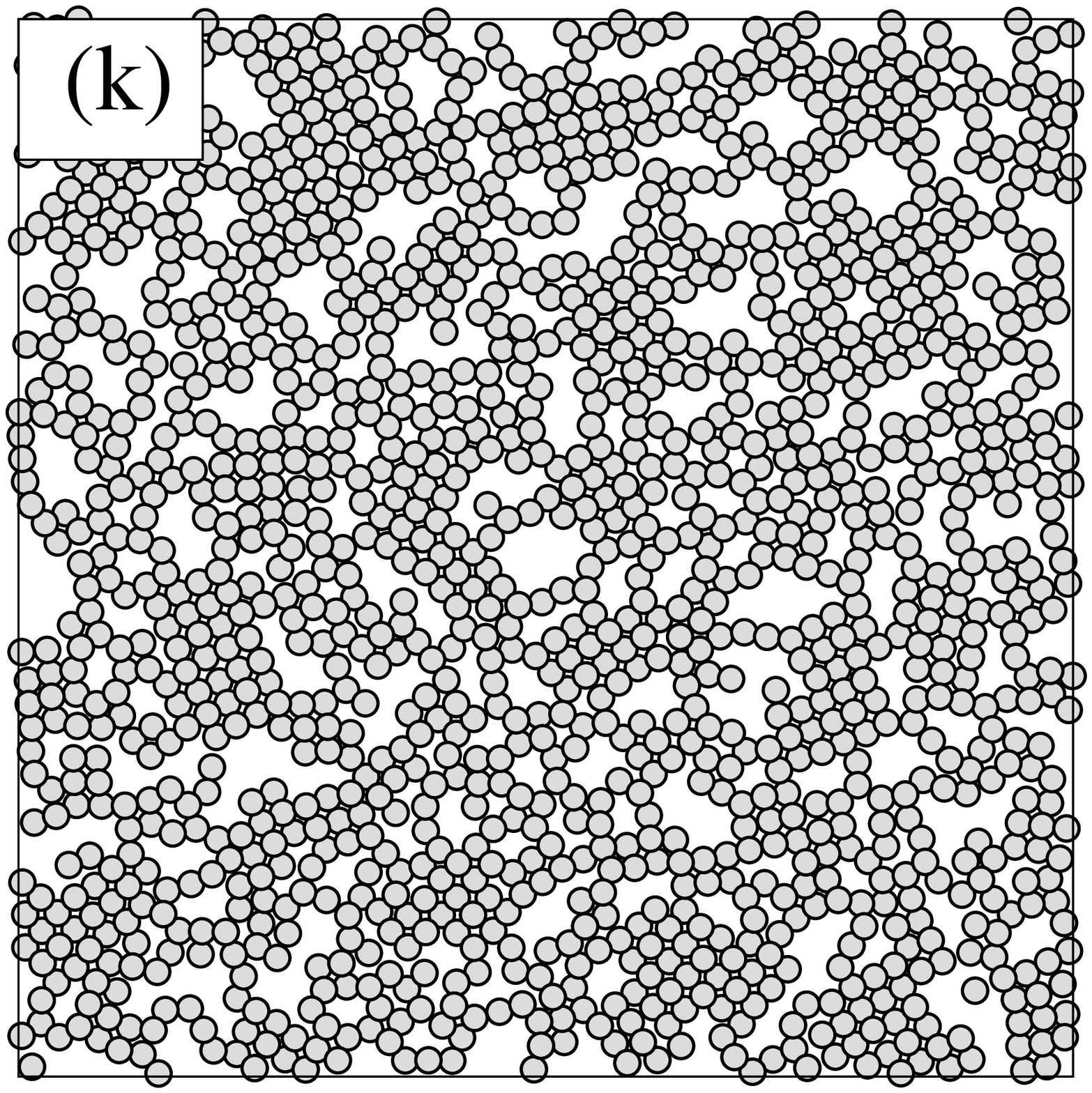}
\label{isochoric-70-temp-23}
}
\subfigure{
\includegraphics[clip=true, trim=0 0 0 0, height=2.6cm, width=2.6cm]{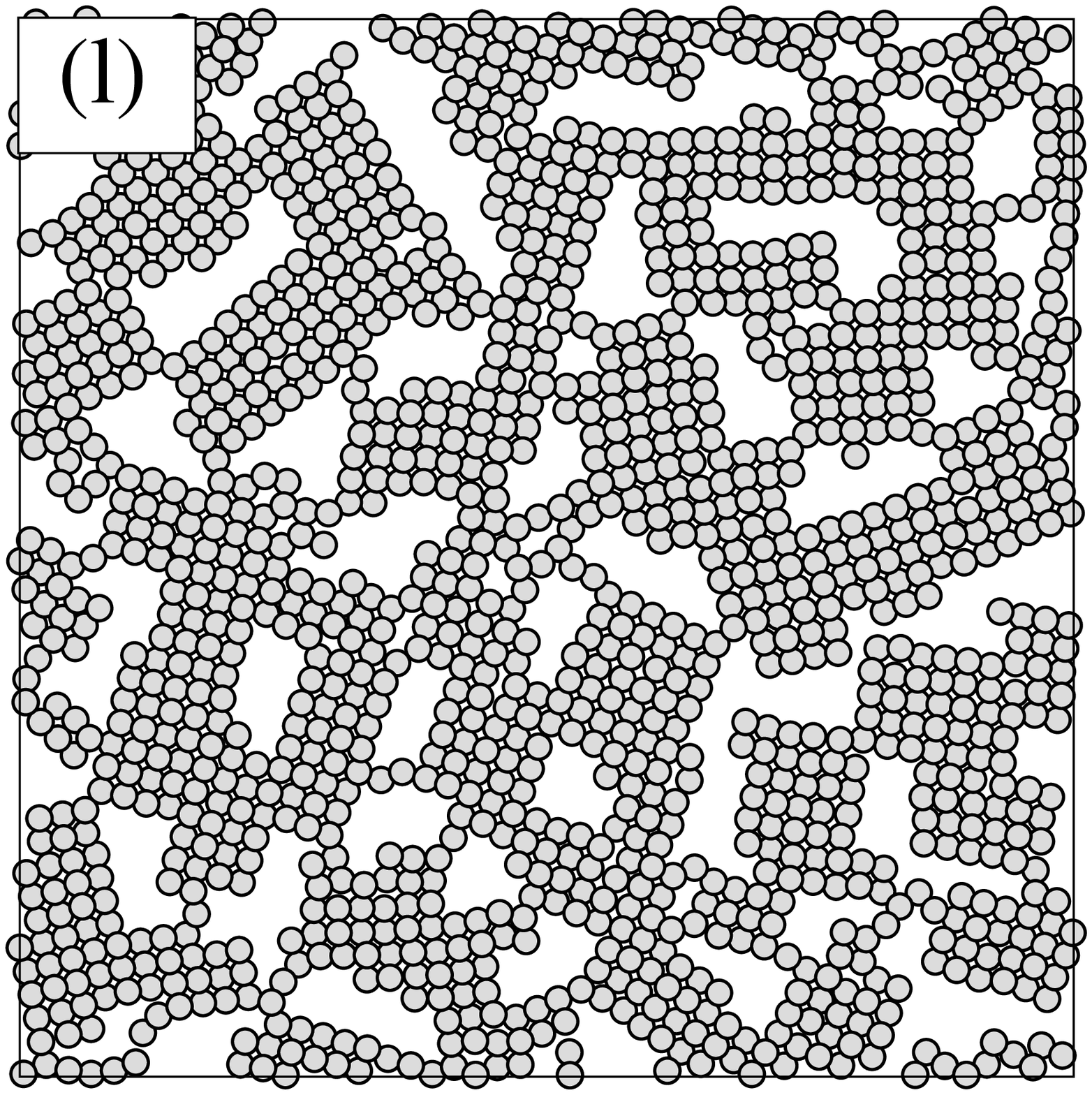}
\label{isochoric-70-temp-18}
}
\caption{Sample configurations for $A_{\phi}$ equal to (a-c) $40\%$, (d-f) $50\%$, (g-i) $60\%$, and (j-l) $70\%$. Configurations are shown for  $T^*$ equal to \subref{isochoric-40-temp-40}\subref{isochoric-50-temp-40}\subref{isochoric-60-temp-40}\subref{isochoric-70-temp-40} 4.0, \subref{isochoric-40-temp-23}\subref{isochoric-50-temp-23}\subref{isochoric-60-temp-23}\subref{isochoric-70-temp-23} 2.3, and \subref{isochoric-40-temp-18}\subref{isochoric-50-temp-18}\subref{isochoric-60-temp-18}\subref{isochoric-70-temp-18} 1.8.}.
\label{isochore-40}
\end{figure}

In Figs.~\ref{isochore-10} and~\ref{isochore-40}, we show sample configurations for different $T$ taken from isochores at $A_\phi=1\% - 70\%$.  At $A_\phi=10\%$, we see that single disks initially cluster into short chains (clusters one particle wide) as $T$ decreases, but then form more compact clusters at the lowest $T$.  We see similar behavior at $A_\phi=40\%$, although the higher density encourages longer, more intertwined chains that thicken as $T$ decreases.  At the lowest $T$,  the system comprises rectangular clusters connected with chains.  

The right columns of Figs.~\ref{isochore-10} and~\ref{isochore-40} show the progression of the system at low $T$ as $A_\phi$ increases from individual crystallites, to crystallites connected by chains, to crystallites with very few chain-like structures.  By $A_\phi=0.70$, there are very few non-crystal-like particles.

What is clear from these pictures is that cluster-cluster repulsion, arising from the cumulative long-range disk-disk repulsions, prevents the formation of a bulk-like crystalline region.  The local crystal-like ordering within clusters induced by lowering $T$ does not result in the appearance of a crystal phase.  It would be difficult to classify the system as being one of phase separated crystal and fluid, despite expecting the system to be fully crystalline at higher densities.

\begin{figure}[ht]
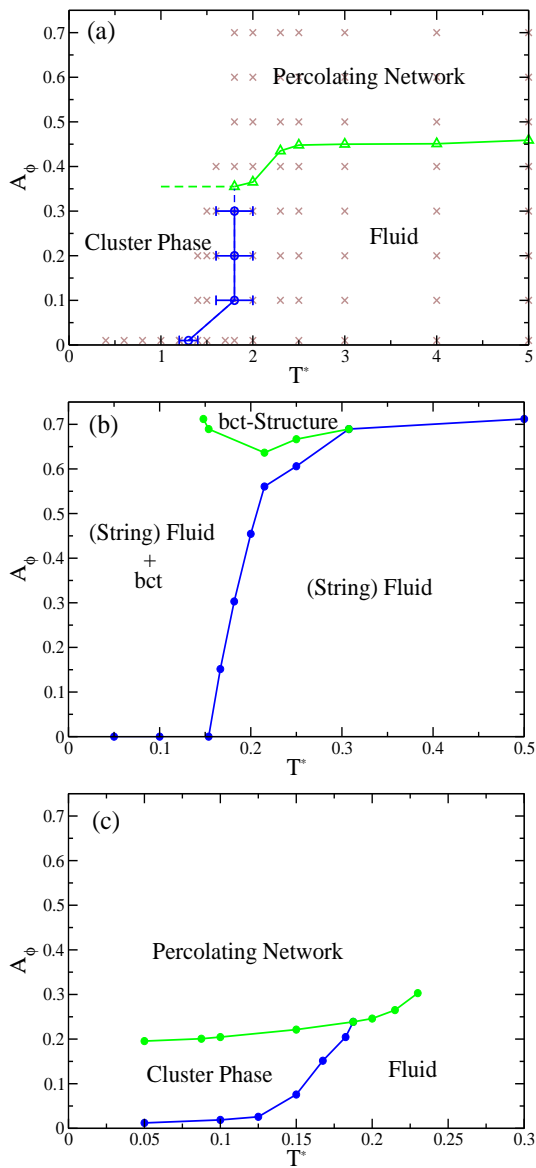

\centering
\subfigure{
\includegraphics[clip=true, trim=0 0 0 0, height=5.0cm, width=7.0cm]{fig7a}
\label{ahmad-phase-diagram} 
}
\subfigure{
\includegraphics[clip=true, trim=0 0 0 0, height=5.0cm, width=7.0cm]{fig7b}
\label{hynninen-phase-diagram} 
}
\subfigure{ 
\includegraphics[clip=true, trim=0 0 0 0, height=5.0cm, width=7.0cm]{fig7c}
\label{sciortino-phase-diagram}
}
\caption{(Color online)
Phase diagrams for comparable systems.
\subref{ahmad-phase-diagram} Our system of chains of dipolar spheres.
\subref{hynninen-phase-diagram} 
The three dimensional system of polarizable spheres in an external field studied by  Hynninen et al.~\cite{dijkstra}. 
We have redrawn their figure after converting packing fraction to area fraction and dipolar strength to reduced temperature.
\subref{sciortino-phase-diagram} Phase diagram as adapted from~\cite{toledano} for a 
colloidal system with short-range depletion attraction and long-range electrostatic repulsion.}
\end{figure}

To categorize the structures seen, we plot the phase diagram for the system in Fig.~\ref{ahmad-phase-diagram}, where we have indicated the percolation line (where we define the percolating cluster by considering bonded particles regardless of local order) and a line where $\Delta F(n)$ first exhibits a minimum at $n > 0$, the ``cluster line''.

\begin{figure}[htp]
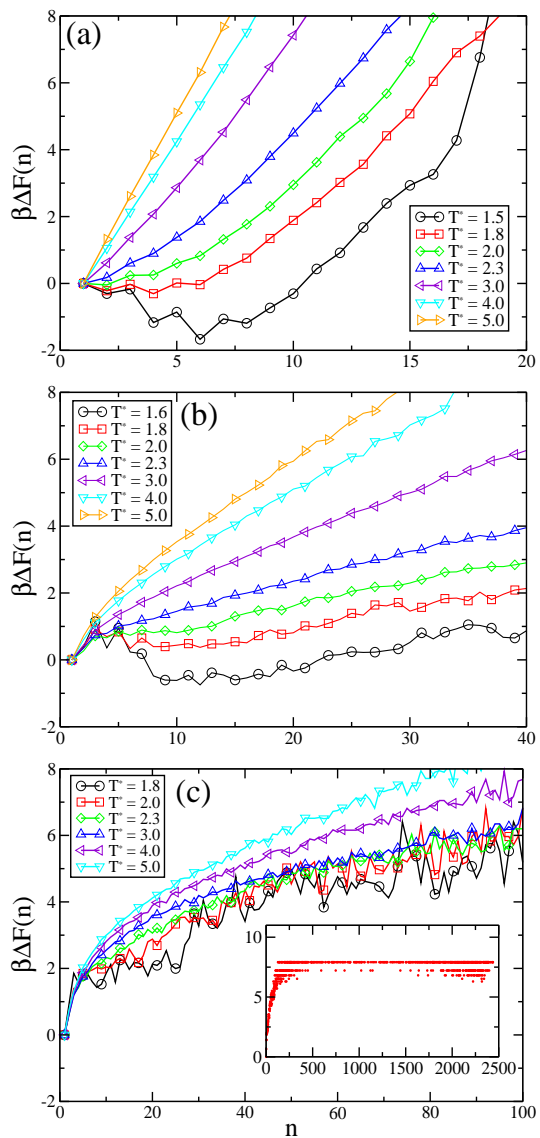

\centering
\subfigure{
\includegraphics[clip=true, trim=0 0 0 0, height=4.8cm, width=7.0cm]{fig8a}
\label{work-10}
}
\subfigure{
\includegraphics[clip=true, trim=0 0 0 0, height=4.8cm, width=7.0cm]{fig8b}
\label{work-30}
}
\subfigure{
\includegraphics[clip=true, trim=0 0 0 0, height=5.0cm, width=7.0cm]{fig8c}
\label{work-40}
}
\caption{(Color online)
The work of forming an $n$-sized cluster regardless of local geometry for $A_{\phi}$  equal to 
\subref{work-10} $10\%$, 
\subref{work-30} $30\%$, and 
\subref{work-40} $40\%$ (inset shows data for all $n$ for $T^*=2.0$).  $\beta=(k_B T)^{-1}$.}
\label{work}
\end{figure}

The $\Delta F(n)$ are shown in Figs.~\ref{work-10}-\ref{work-40} for area fractions well within and near the percolation threshold.  
For $A_\phi=10\%$, a minimum in $\Delta F(n)$ for small but finite $n$ begins to develop at $T^*\approx 1.8$, moving to larger $n$ as $T$ decreases.  The transition from the high $T$ fluid, characterized by stringy clusters, to the low $T$ cluster phase, characterized by more compact, squarish clusters, appears to be continuous with no free energy barrier between monomers and larger cluster at low $T$.
For $A_\phi=30\%$, the picture is similar, except that the slope of $\Delta F(n)$ is smaller in the cluster regime, and that there appears to be a small barrier on the order of $k_B T$ separating monomers from larger clusters.  While this would lend a first-order flavor to the transition, the minimum in $\Delta F(n)$ is local and $\Delta F(n)$ increases monotonically beyond it.
For $A_\phi=40\%$, which crosses the percolation threshold as $T$ decreases, no local minimum in $\Delta F(n)$ forms for small $n$.  Rather, accompanying percolation, a local minimum develops at $n$ on the order of the system size.

\begin{figure}[htp]
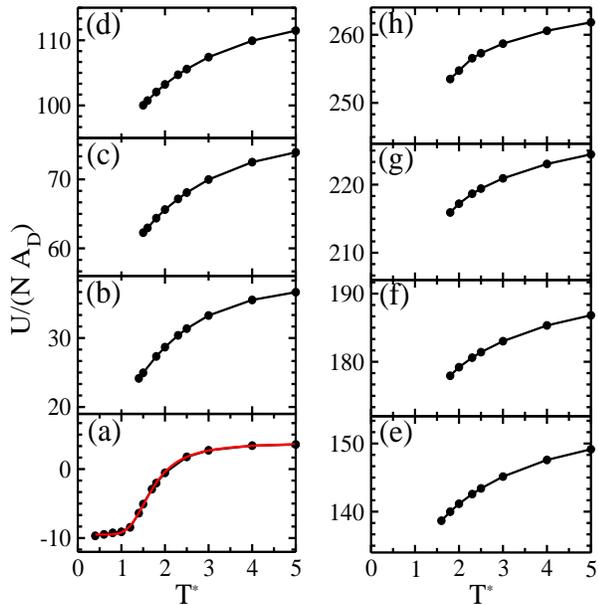

\centering
\subfigure{
\includegraphics[clip=true, trim=0 0 0 0, height=8.0cm, width=3.9cm]{fig9a}
\label{energy-comparing1}
}
\subfigure{
\includegraphics[clip=true, trim=0 0 0 0, height=8.0cm, width=3.6cm]{fig9b}
\label{energy-comparing2}
}
\caption{(Color online)
Potential energy per particle as a function of $T^*$ for $A_\phi$ equal to  
(a) $1\%$ (red curve indicates two-state model fit), (b) $10\%$, (c) $20\%$, (d) $30\%$, (e) $40\%$, (f) $50\%$, (g) $60\%$, (h) $70\%$.}
\label{energy-comparing}
\end{figure}

The potential energy as a function of $T$ for the isochores is plotted in Fig.~\ref{energy-comparing}. For $A_\phi=1\%$, where we approach the clustered ground state, the potential energy is well described by a two-state model, with the exception at low $T$ where the vibrational contribution of $3/2 k_B T$ is not accounted for in the model.  For $A_\phi = 10\%$ and $20\%$ our lowest $T$ data appear to be just at or below the inflection point in $U(T)$, while for $A_\phi \ge 30\%$, we have not yet located the peak in the heat capacity. Thus, we do appear to have a heat capacity peak accompanying the change demarcated by the appearance of an $n > 0$ minimum in $\Delta F(n)$, at least at low $A_\phi$.

\begin{figure}[ht]
\centering
\includegraphics[clip=true, trim=0 0 0 0, height=4.8cm, width=7.1cm]{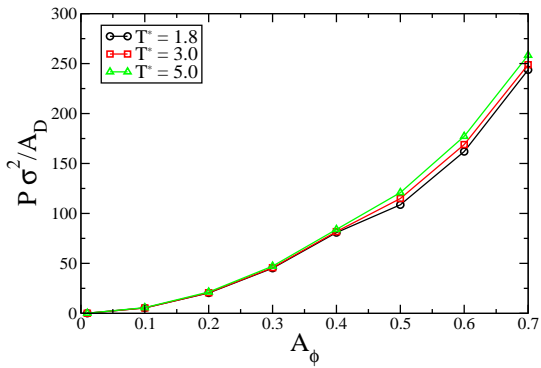}
\caption{(Color online) Pressure isotherms. The inflection coincides with percolation.}
\label{pressure-figure}
\end{figure}

\begin{figure}[ht]
\centering
\includegraphics[height=4.8cm, width=7.1cm]{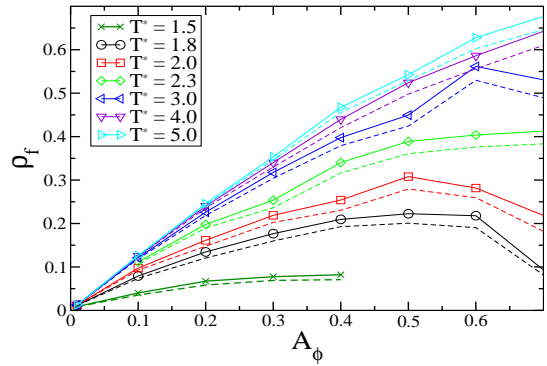}
\caption{(Color online) Reduced number density of the liquid-like portion of the system $(\rho_{f})$ as a function of $A_{\phi}$ for different isotherms at 
$\alpha=103^\circ$ (solid lines) and $\alpha=115^o$ (dashed lines).}
\label{isotherm-liquid-particles}
\end{figure}

Pressure isotherms, as obtained through Eq.~\ref{pressure} are plotted in Fig.~\ref{pressure-figure}.  At packing fractions near the percolation transition, the pressure shows an inflection, indicating a compressibility maximum, that grows with decreasing $T$.  It is not surprising that extrema in response functions appear together with percolation~\cite{skvor}.

Fig.~\ref{isotherm-liquid-particles} shows the packing fraction of the non-locally-crystal-like, i.e.~liquid-like, portion of the system.  
As discussed in the previous section, there is some ambiguity in discerning liquid-like particles from crystal-like particles, 
particularly for particles connecting crystallites.  As such, we plot curves for two values of $\alpha$ (an angle that determines whether particles bridging crystal-like domains are crystal-like themselves), and find only a quantitative difference in the curves.
The $\rho_f$ isotherms become flatter as $T$ decreases, which is consistent with an interpretation of our cluster line as (a remnant of) the phase boundary between the fluid and the solid-fluid coexistence phase.
At $T$ well below this line, $\rho_f(A_\phi)$ is quite flat.

\begin{figure}[htp]
\centering
\subfigure{
\includegraphics[clip=true, trim=0 0 0 0, height=4.8cm, width=7.0cm]{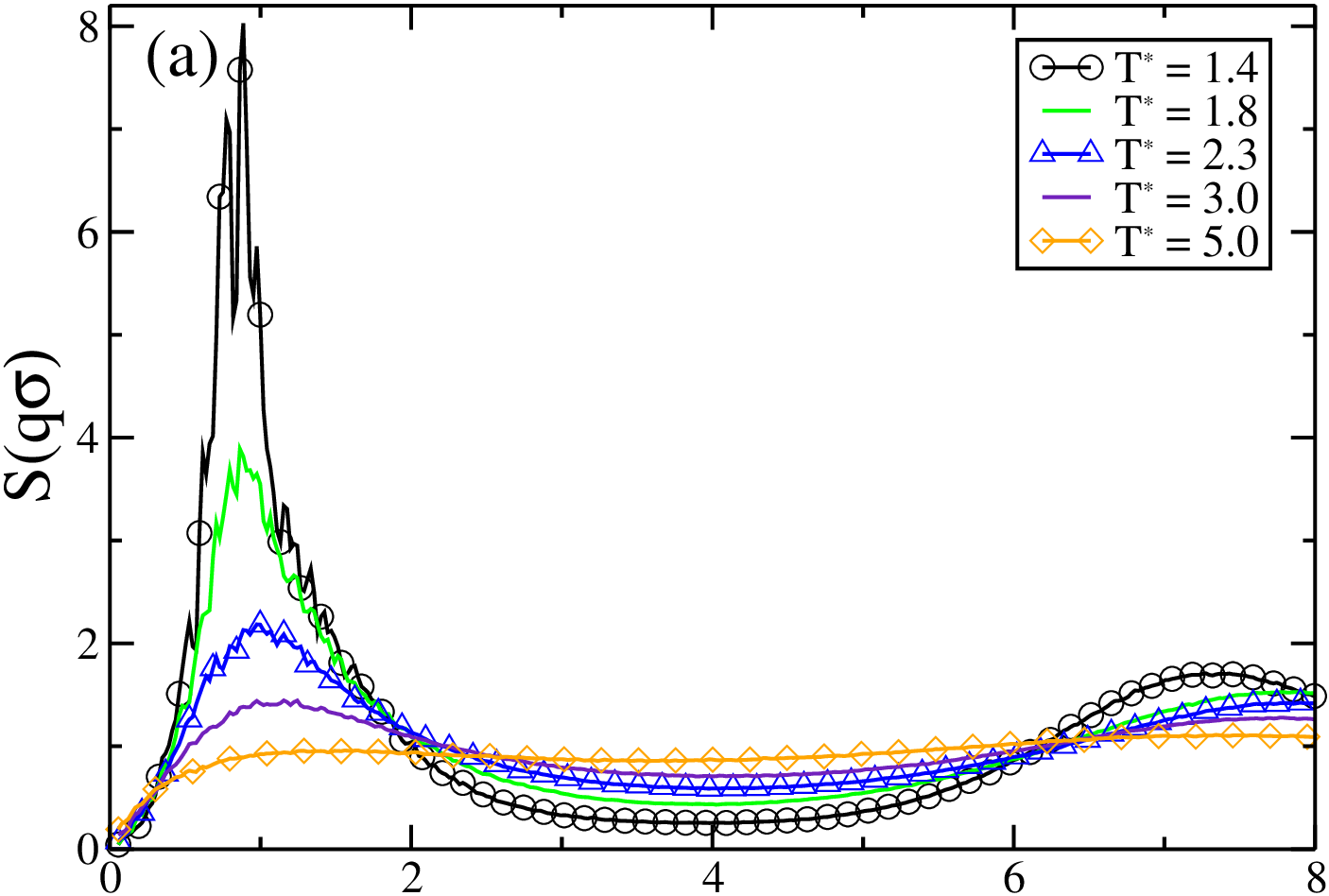}
\label{isochoric-stru-10}
}
\subfigure{
\includegraphics[clip=true, trim=0 0 0 0, height=4.8cm, width=7.0cm]{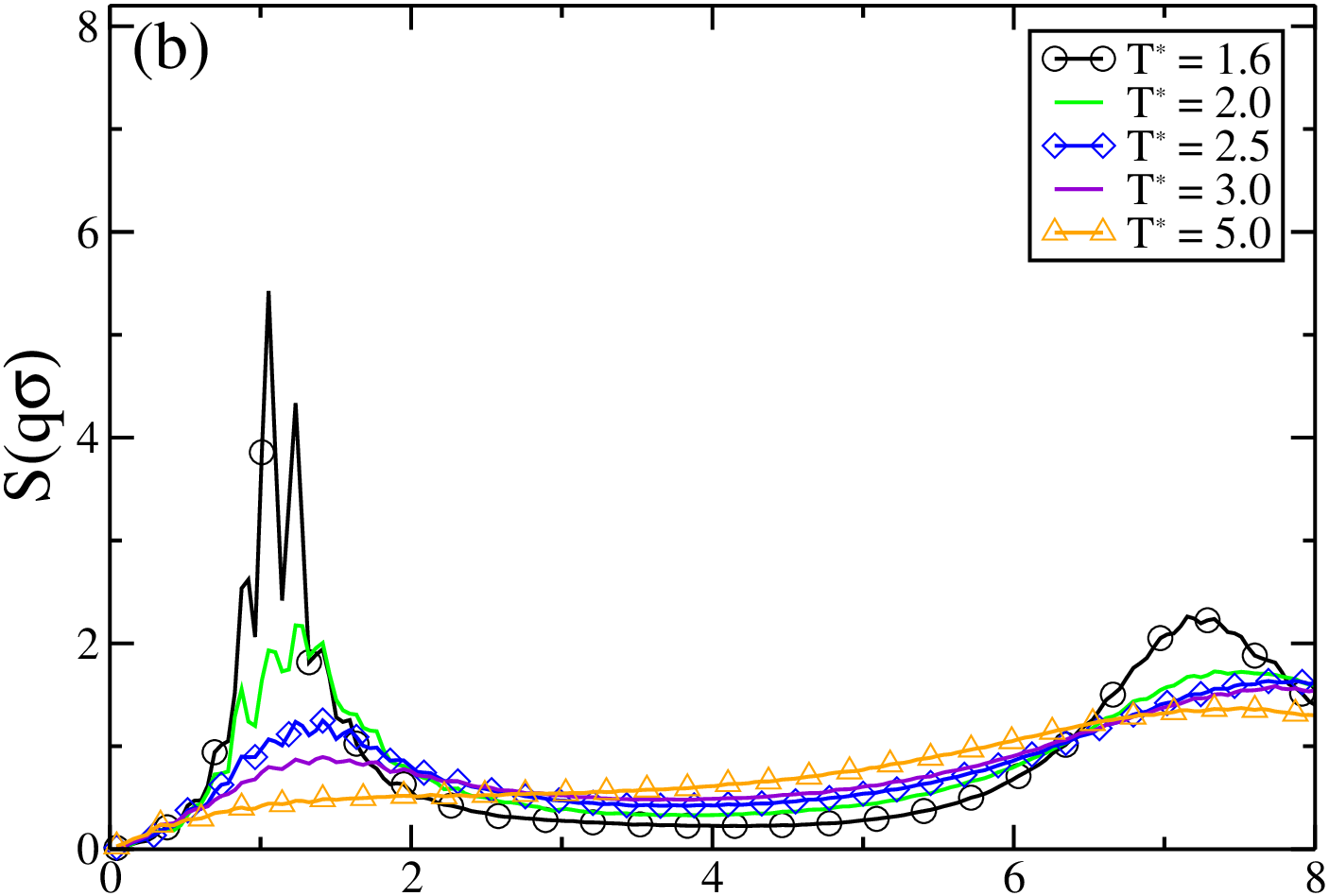}
\label{isochoric-stru-40}
}
\subfigure{
\includegraphics[clip=true, trim=0 0 0 0, height=5.0cm, width=7.0cm]{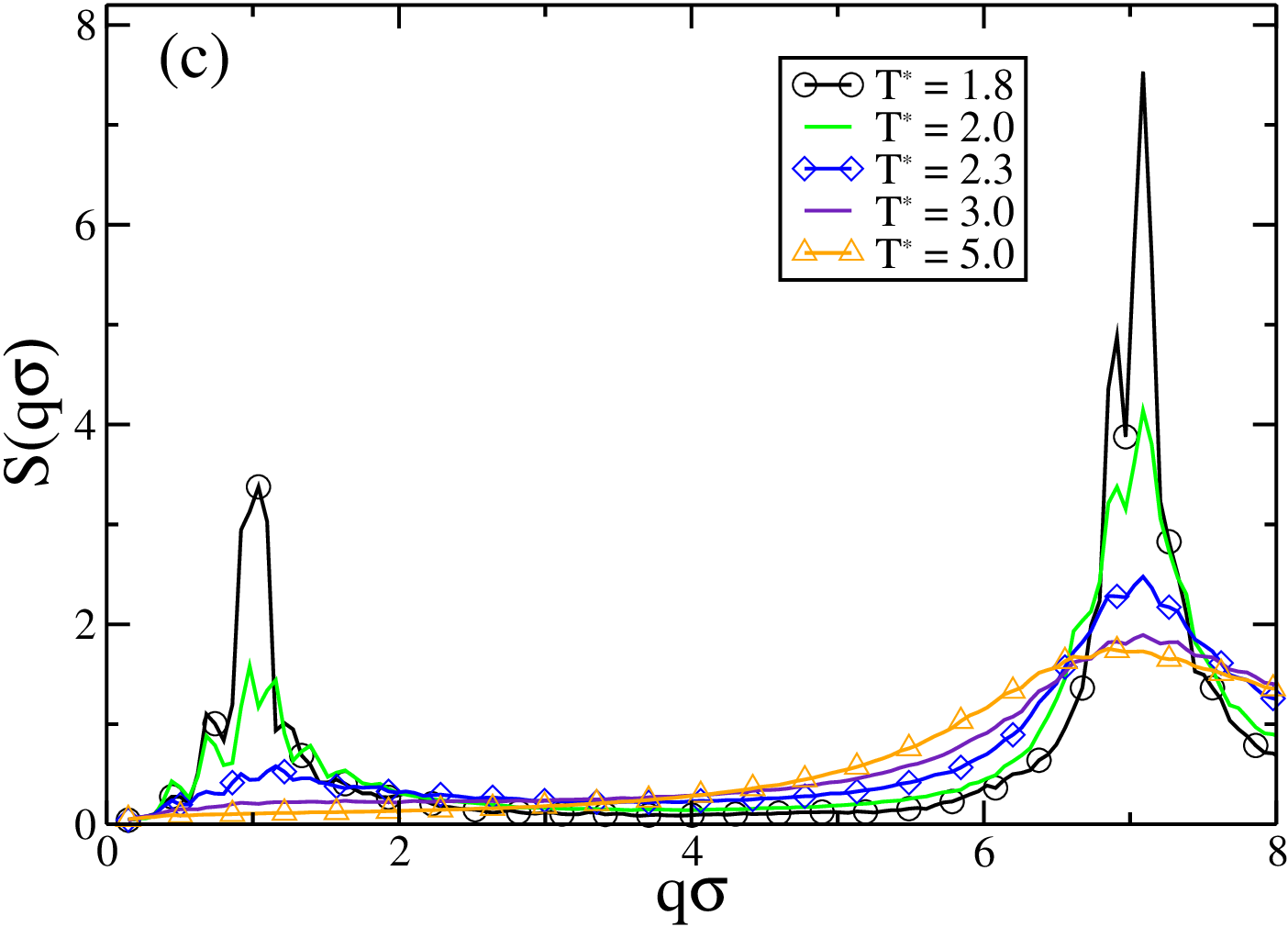}
\label{isochoric-stru-70}
}
\caption{(Color online) The structure factor $S(q)$ for $A_{\phi}$ equal to  \subref{isochoric-stru-10} 
$10\%$, \subref{isochoric-stru-40} $40\%$, \subref{isochoric-stru-70} $70\%$.}
\label{isochoric-stru}
\end{figure}

In Fig.~\ref{isochoric-stru}, we plot $S(q)$ for $A_\phi=10\%$, 40\% and 70\%.  In all cases, ordering on the scale of nearest neighbors is closely followed by clustering as evidenced by the peak near $q\sigma=1$.  The location of the cluster peak appears not to change much with $A_\phi$, indicating that the distance between cluster centers does vary a great deal with density.  The high value of the cluster peak at $A_\phi=10\%$ may at first seem to indicate a cluster crystal since in two dimensions a peak height greater than four is characteristic of a crystal~\cite{gasser}.  However, to draw any such conclusion, it would be necessary to divide $S(q)$ by the number of particles in a cluster, which is approximately four, yielding an effective peak height of two.  Visually, we do not see ordering of the crystallites.
On the other hand, for $A_\phi=70\%$, $S(q)$ reaches a height of four at $q$ corresponding to nearest neighbors, nominally signifying crystallization.

\begin{figure}[htp]
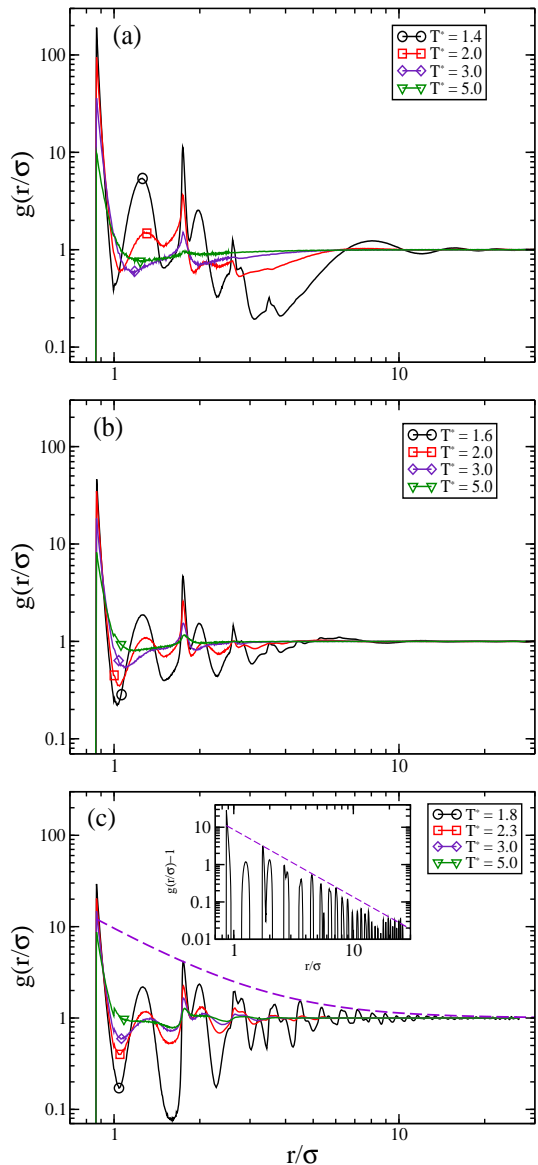

\centering
\subfigure{
\includegraphics[clip=true, trim=0 0 0 0, height=5.0cm, width=7.0cm]{fig13a}
\label{isochoric-pair-10}
}
\subfigure{
\includegraphics[clip=true, trim=0 0 0 0, height=5.0cm, width=7.0cm]{fig13b}
\label{isochoric-pair-40}
}
\subfigure{
\includegraphics[clip=true, trim=0 0 0 0, height=5.0cm, width=7.0cm]{fig13c}
\label{isochoric-pair-70}
}
\caption{(Color online) The pair correlation function $g(r)$  for $A_{\phi}$ equal to  \subref{isochoric-pair-10} $10\%$, 
\subref{isochoric-pair-40} $40\%$ 
and \subref{isochoric-pair-70} $70\%$, where the dashed curve is $8.6 (\sigma/r)^{1.8}+1$. Inset shows
$g(r)-1$ for $T^*=1.8$ and $8.6 (\sigma/r)^{1.8}$ . }
\label{pair-correlation}
\end{figure}

\begin{figure}
\centering
\includegraphics[height=5.0cm, width=7.0cm]{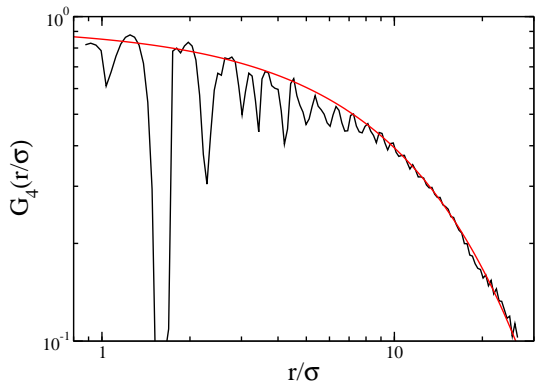}
\caption{(Color online) $G_{4}(r)$ as a function of $r$ at $A_{\phi} = 70\%$ and $T = 1.8$.  Red curve shows 
an exponential fit $A\exp{[-r/\xi]}$, with $A=0.928$ and $\xi = 11.67 \sigma$.}
\label{g4}
\end{figure}

In Fig.~\ref{pair-correlation} we plot $g(r)$ for $A_{\phi}=10\%$, $40\%$ and $70\%$, illustrating the $T$ evolution below, near and above percolation. At $10\%$, as $T$ decreases, we see the emergence of neighbor peaks at short range, while a long range peak emerges near $8\sigma$ only once the cluster line is crossed.  For  $A_{\phi}=70\%$, near $T^*=2.0$ and below, the neighbor peaks extend to longer distances.  The peak heights appear to decrease as a power law, as highlighted by a power law curve with exponent $-1.8$ drawn as a guide to the eye in Fig.~\ref{pair-correlation}(c).  Similar, but less pronounced behavior is evident at  $A_{\phi}=60\%$ and $50\%$ (not shown).  Fig.~\ref{pair-correlation}(b) shows  $A_{\phi}=40\%$ exhibiting cross-over behavior with a cluster peak emerging near $r=6\sigma$at low $T$.
In Fig.~\ref{g4} we show $G_4(r)$ for $A_{\phi}=70\%$ and $T=1.8$, the state point showing the longest range translational order through $g(r)$.  However, the orientational order quantified by $G_4(r)$ appears to decay exponentially, as shown by the curve $A\exp{[-r/\xi]}$, with $A=0.928$ and $\xi = 11.67 \sigma$,
overlaid on the data.

\section{Discussion}

This study was motivated by the desire to better understand the phase behavior of a system of polarizable colloids under the influence of an alternating, external field.  The system has been hitherto studied as a bulk system of dipolar hard sphere particles with all dipole moments aligned along one axis.  With our reduction of the system to one of chains of finite length, we make explicit the  emergent short range attraction and long range repulsion between the constituents of what can be effectively thought of as a two dimensional system.  This type of interaction generically promotes finite clustering and suppresses the formation of a bulk phase that would otherwise arise in a fluid-solid coexistence.  

The repulsive part of our potential becomes weaker with respect to the short range attraction as chain length increases, disappearing in the infinite chain length limit.  One can readily confirm that the interaction between two parallel lines of uniform and aligned polarization decays to zero as the length of the chains increases.  In this limit of long chains, the only source of interaction (attraction) is the discrete nature of the spherical dipolar spheres constituting the chains.  Thus, one can  see why for a large, three dimensional system, bulk fluid-crystal phase coexistence emerges.

For our system at densities below percolation, the high $T$ phase is characterized by a distribution of clusters sizes, where monomers are the most common state and clusters tend to be string-like.  As $T$ decreases, the distribution shifts so that the most common cluster size is finite, and clusters tend to be more compact.  The clusters themselves do not order into a lattice at the $T$ studied.  At very low density, we expect the transition to the cluster phase to continue to occur at progressively lower $T$ as the energetics of clustering competes with an entropically favored monomeric state.
At densities above percolation, the homogeneous state coalesces into a system of connected crystallites, which increase in size with increasing density.

Snapshots of configurations are qualitatively quite similar to experimental ones~\cite{ay2002}, and our results confirm that even
for uncharged colloids, dipolar repulsions between chains stabilize clusters and coarsening (to the bulk) should not be expected beyond some finite  
equilibrium size.

Having suppressed bulk phase separation, we obtain a qualitatively different phase behavior from that obtained for the three dimensional study 
by Hynninen~\cite{dijkstra} and coworkers.  For comparison, we present an adaptation of their phase diagram in Fig.~\ref{hynninen-phase-diagram}, after converting their packing fraction to area fraction $(A_D = 3/2 \eta)$, where $\eta$ is the packing fraction.  Our $T$ scale is rather different, owing to our precondition that our system comprises fully formed and unchanging chains.  Despite the lack of full phase separation, remnants of first order character appear in our cluster phase, namely, a possibly non-zero free energy barrier separating the monomeric state and stable clusters of a larger size, as well as the flatness of the $\rho_f(A_\phi)$ curve.
As mentioned in the Introduction, it would be interesting to study the approach of our two dimensional model to the three dimensional case by progressively increasing $L$.

Our phase behavior appears to be closer in spirit to that of a three dimensional system of particles interacting through isotropic competing short range attraction and long range repulsion studied by Toledano {\it et al} \cite{toledano}.  In Fig.~\ref{sciortino-phase-diagram} we show their phase diagram, again rescaling packing fraction.  One point to emphasize for our system is the importance of percolation, for which we see an accompanying inflection in the pressure.

At densities above percolation, we see what appears to be the remnant of bulk crystallization in the formation of connected crystallites below a $T$ similar to the $T$ at which the (unpercolated) cluster phase appears at lower density.  While a large nearest neighbor peak in S(q) indicates crystallinity, the orientational and translational order do not extend in space appreciably beyond the characteristic size of crystallites. Interestingly, the peaks in $g(r)$ decay with a power law dependence, though with a larger exponent ($\sim1.8$) than what is typical for orientational correlation function decay in a hexatic phase~\cite{gasser}.

Even though the crystallites in the percolated regime do not seem to align to form a structure with long range order, and this may be understood in terms of the degeneracy of the long range energy between two clusters with respect to precise orientation, at sufficiently high density the system must become fully crystalline.  Presumably, the size of the crystallites continues to increase with density, perhaps forming an ordered phase dotted with voids, but the nature of the transition is for us an open question.

The transition to the crystal would perhaps be better rigorously studied through another model.  One drawback of the present model is that it is difficult to handle the long-range nature of the interaction in a precise way, and the dynamics become quite slow at high density.  Our potential, which we determine numerically, converges too slowly to  $\sim r^{-3}$ for us to be able to  take advantage of Ewald techniques.  Perhaps the formation of the crystal phase could be studied with simpler potentials, or even potentials with finite repulsive tails, as were studied recently in Ref.~\cite{haw2010}.  Such a potential would also simplify a more thorough study of the thermodynamic behavior near percolation.

Finally, we find no evidence at very low density of the emergence of the cellular structure of the ``void phase'' of Ref.~\cite{amit}.  Clearly, the physics of dipolar hard spheres assembled into chains   monodispersed in length is not solely responsible for the appearance of this phase.  Perhaps the formation of chains helps stabilize structures that begin to form as a result of hydrodynamic forces, but this is mere speculation.

\section{Conclusions}

In this work, we reduce a model of aligned dipolar hard spheres to one of interacting finite chains of particles, treating the chains as a collection of disks in two dimensions.  The long range repulsion between disks (chains) promotes the appearance of a cluster phase at the expense of bulk crystal formation.  Percolation at higher density is accompanied by a pressure anomaly.  The percolated phase at low $T$ is characterized by connected crystallites exhibiting orientational disorder that increase in size with increasing density and power-law decay of $g(r)$.  The description of the transition to the crystal phase at densities higher than those presented here is of great interest to us, but such a study would be better done with a more computationally tractable potential.  We find no evidence for the appearance of the ``void'' phase found experimentally at very low density.

\section*{Acknowledgments}
We thank ACEnet for computing resources and research funding, as well NSERC for funding.  We thank Anand Yethiraj for many useful discussions.

\end{document}